\documentclass[12pt]{article}
\usepackage{latexsym}
\usepackage{mathrsfs}
\usepackage{amsmath}
\usepackage{amssymb}
\usepackage{wasysym}
\newcommand{\half}{\frac{\scriptstyle 1}{\scriptstyle 2}}
\newcommand{\C}{\mathbb{C}}

\newcommand{\CP}{\mathbb{CP}}
\newcommand{\RP}{\mathbb{RP}}
\newcommand{\PT}{\mathbb{PT}}
\newcommand{\R}{\mathbb{R}}
\renewcommand{\P}{\mathbb{P}}
\newcommand{\PS}{\mathbb{PS}}
\newcommand{\bbS}{\mathbb{S}}

\newcommand{\F}{\mathscr{F}}
\newcommand{\cH}{\mathcal{H}}
\newcommand{\scri}{\mathscr{I}}
\newcommand{\M}{\mathbb{M}}
\newcommand{\CM}{\mathbb{CM}}

\newcommand{\T}{\mathbb{T}}
\newcommand{\Z}{\mathbb{Z}}
\newcommand{\cD}{\mathcal{D}}
\newcommand{\p}{\partial}
\renewcommand{\d}{\mathrm{d}}
\newcommand{\e}{\mathrm{e}}

\newcommand{\D}{\mathrm{D}}

\newcommand{\Oc}{\mathcal{O}}
\newcommand{\cS}{\mathcal{S}}
\newcommand{\Aut}{\mathrm{Aut}}
\newcommand{\SL}{\mathrm{SL}}
\newcommand{\PSL}{\mathrm{PSL}}

\newcommand{\Herm}{\mathrm{Herm}}
\newcommand{\SU}{\, \mathrm{SU}}
\newcommand{\U}{\, \mathrm{U}}
\newcommand{\su}{\, \mathrm{su}}
\newcommand{\PSU}{\, \mathrm{PSU}}
\newcommand{\SO}{\, \mathrm{SO}}
\newcommand{\PSO}{\, \mathrm{PSO}}
\newcommand{\GL}{\mathrm{GL}}

\newcommand{\diag}{\, \mathrm{diag}}
\newcommand{\tr}{\, \mathrm{tr}}
\newcommand{\bx}{{{\mathbf{x}}}}
\newcommand{\by}{{{\mathbf{y}}}}

\newcommand{\be}{\begin{equation}\label}
\newcommand{\ee}{\end{equation}}
\newcommand{\bea}{\begin{eqnarray}\label}
\newcommand{\eea}{\end{eqnarray}}
\newcommand{\proof}{ \noindent {\bf Proof:} }

\newtheorem{defn}{Definition}[section]
\newtheorem{thm}{Theorem}

\newtheorem{propn}{Proposition}[section]

\newtheorem{corol}{Corollary}[section]
\newtheorem{lemma}{Lemma}[section]
\newcommand{\rmk}{\noindent {\bf Remark: }}

\begin{document}
\title{Global anti-self-dual Yang-Mills fields in split signature and
 their scattering} \author{L.J.Mason \\ \small{ \it{ The Mathematical
 Institute, 24-29 St Giles, Oxford OX1 3LB, England}}}

\maketitle

\abstract{This article concerns solutions to the anti-self-dual Yang
  Mills (ASDYM) equations in split signature that are global on the
  double cover of the appropriate conformally compactified Minkowski
  space $\widetilde\M=S^2\times S^2$.  Ward's ASDYM twistor
  construction is adapted to this geometry using a correspondence
  between points of $\widetilde\M$ and holomorphic discs in $\CP^3$,
  twistor space, with boundary on the real slice $\RP^3$.  Smooth
  global $\U(n)$ solutions to the ASDYM equations on $\widetilde\M$
  are shown to be in 1:1 correspondence with pairs consisting of an
  arbitrary holomorphic vector bundle $E$ over $\CP^3$ together with a
  positive definite hermitian metric $H$ on $E|_{\RP^3}$.  There are
  no topological or other restrictions on the bundle $E$.  In
  ultrahyperbolic signature solutions are generically non-analytic or
  only finitely differentiable and such solutions arise from a
  corresponding choice of regularity for $H$.  When $E$ is trivial,
  the twistor data consists of the Hermitian matrix function $H$ on
  $\RP^3$ up to constants and the correspondence provides a nonlinear
  generalisation of the X-ray transform. In general it provides a
  higher-dimensional analogue of the (inverse) scattering transform in
  which $H$ plays the role of the reflection coefficient and $E$ the
  algebraic data.
  
  Explicit examples are constructed for different choices of the
  topology of $E$.
  
  A scattering problem for ASDYM fields on affine Minkowski space in
  split signature is set up and it is shown that sufficiently small
  data at past null infinity uniquely determines data at future null
  infinity by taking a family of holonomies associated to the initial
  data followed by a sequence of two Birkhoff factorizations.  The
  scattering map is simple at the level of the holonomies, but
  non-trivial at the level of the connection in the non-abelian case.}

\section{Introduction}
The anti-self-dual Yang-Mills (ASDYM) equations have long been known
to be an integrable system, Ward (1977), Belavin \& Zakharov (1978).
However, they only admit real solutions in Euclidean (positive
definite) signature or split (ultrahyperbolic) signature.  The
integrability has allowed a substantial study of solutions to the
equations in Euclidean signature, see for example Atiyah (1979).  Due
to the unphysical and peculiar nature of ultrahyperbolic differential
equations, the solutions in split signature have not been so much
studied.  Nevertheless, the ASDYM equations in split signature have
importance because their symmetry reductions give rise to a wide class
of integrable evolution equations, see Ablowitz \& Clarkson (1991) and
Mason \& Woodhouse (1996) for surveys.  Although the main motivation
for this work is mathematical, it is worth noting also that the recent
work of Witten (2004) and others has shown that split signature
versions of the twistor correspondence provide a useful calculus for
twistor string theory with significant applications in perturbative
gauge theory.\footnote{however, the use of the `wrong
signature' has also perhaps led to some of the technical problems of
twistor-string theory.}

The purpose of this paper is to adapt the Ward transform to split
signature in such a way as to provide a full study of global solutions
to the ASDYM equations on the double cover of the conformal
compactification of Minkowski space and also of the classical
scattering of such fields on affine Minkowski space.  The mathematical
motivation for the study of the global problem in split signature
arises from two sources.  The abelian version of the Ward
correspondence extends to all massless fields as the Penrose transform
between analytic first cohomology classes on regions in twistor space
$\PT=\CP^3$ and linear massless fields on corresponding regions in
space-time.  In split signature, twistor space has a naturally defined
real slice $\PT_\R=\RP^3$ such that conformally compactified Minkowski
space $\M$ is the space of real lines in $\RP^3$ (the real Klein
correspondence).  There is another transform in split signature, the
(generalized) X-ray transform: a smooth function $f$ (or section of an
appropriate line bundle) on the real slice $\PT_\R=\RP^3\subset\PT$
can be integrated along lines in $\PT_\R$ to yield a function $\phi$
on ultrahyperbolic Minkowski space.  It is a classical result that
$\phi$ is a solution to the ultrahyperbolic wave equation and that all
such solutions determine a unique $f$ on $\RP^3$, John (1938).  There
is a particular puzzle in that, globally on $\CP^3$, the appropriate
cohomology group is finite dimensional whereas the X-ray transform
shows that there are an infinite dimensional family of solutions.  One
can nevertheless naively think of the function $f$ as a preferred Cech
cocycle, Atiyah (1979).  However, the task of finding how this cocycle
comes to be preferred and more generally what the precise relationship
is between the Penrose and the X-ray transform has led to a
substantial literature: see for example Guillemin \& Sternberg (1986),
Woodhouse (1992), Mason (1995), Sparling (1998), Bailey, Eastwood,
Gover and Mason (1999, 2003), Bailey \& Eastwood (2001) and remarks in
\S\ref{abelian} of this paper.  Furthermore, it is this X-ray analogue
of the Penrose transform that is predominantly used in Witten (2004).
The first motivation then is to find the appropriate non-linear
extension of the X-ray transform for ASDYM fields.

The second motivation arises from the theory of integrable systems.
For a hyperbolic or parabolic integrable system, the scattering
transform usually expresses general data for a solution in terms of a
combination of solitonic and radiative or dispersive modes, see for
example Faddeev and Takhtajan (1987), Ablowitz and Clarkson (1991).
The solitonic modes are usually described by algebreo-geometric data,
whereas the radiative/dispersive modes are usually described by smooth
functions.
%(Global solutions to the corresponding Euclidean signature elliptic
%equations will generally consist only of the soliton-like solutions,
%which in the case of the anti-self-dual Yang-Mills (ASDYM) equations
%in euclidean signature are the well known instanton solutions.)  
The second motivation then is to find the analogous description of the
ASDYM equations in split signature.  Since many parabolic and
hyperbolic integrable systems are symmetry reductions of the ASDYM
equations in split signature, this would give some general insight as
to how such constructions arise from their twistor descriptions.

The main theorem is a correspondence for solutions to the
anti-self-dual Yang-Mills (ASDYM) equations with compact gauge group
$G$ on the double cover of the conformal compactification of
ultrahyperbolic Minkowski space $\widetilde\M=S^2\times S^2$.  The
correspondence is with certain data on 
complex projective three space $\CP^3$, the twistor space which we denote
by $\PT$, and its real slice $\RP^3$ denoted $\PT_\R$.  We have
\begin{thm}\label{mainthm2} Gauge equivalence classes of $C^\infty$
  solutions to the ASDYM equations on $\widetilde\M$ are in 1-1
  correspondence with principal $G_\C$ bundles $P\rightarrow \PT$
  together with a $C^\infty$ reduction of the structure group to $G$
  over $\PT_\R$ where $G_\C$ is the complexification of $G$. The
  reduction of the structure group over $\PT_\R$ can be expressed as a
  section $H:\PT_\R\rightarrow P/G$ of the $G_\C/G$ bundle $P/G$.
\end{thm}
In the $\U(n)$ case, $P$ is the
principal bundle associated to a holomorphic vector bundle $E$, and
$H$ defines a hermitian metric on the fibres of $E$ restricted to
$\PT_\R$ (when $G=\U(n)$, $G_\C/G$ is the space $\Herm_n$ of $n\times
n$ positive definite hermitian matrices).

The techniques extend to the non smooth case: $C^{k,\alpha}$ solutions
to the ASDYM equations arise when $H$ is in $C^{k+1,\alpha'}$ for
  $\alpha'>\alpha$.  However, the techniques of the proof lose too
  many derivatives in the forward direction to give a definitive
  theorem whereas to obtain a definitive result one would need to gain
  at least one.

The theorem gives the space $\cS_G$ of gauge equivalence classes of
solutions to the ASDYM equations on $\widetilde \M$ the structure of a
fibre bundle $\pi:\cS_G\rightarrow \cH_{G_\C}$ where $\cH_{G_\C}$ is
the space of holomorphic $G_\C$ bundles over $\PT$ and the fibre
$\pi^{-1}(P)$ at $P\in\cH_{G_\C}$ is the space of sections
$H:\PT_\R\rightarrow P/G$.  The space $\cH_{G_\C}$ has many components
labelled by the possible topological types of $P$.  The different
components are not, however, smooth manifolds as arbitrary holomorphic
bundles are allowed, including unstable ones.  Ignoring these
subtleties, the individual connected components of $\cH_{G_\C}$ are
finite dimensional.  However, at a given $P\in \cH_{G_\C}$, the fibre
$\pi^{-1}(P)=\Gamma(\PT_\R,P/G) $ is infinite dimensional being a
twisted analogue of smooth maps from $\RP^3$ to $G_\C/G$ modulo at
worst a finite-dimensional equivalence under the global automorphisms
$\Aut(P)$ of $P$.  When $P$ or $E$ is trivial $\Aut(P)=G_\C$, the
corresponding component of $\cH_{G_\C}$ is a point and so the fibre is
the quotient of the space of smooth maps $\{\mbox{Maps:
}\PT_\R\rightarrow G_\C/G\}$ divided on the left by $G_\C$.  Thus, for
trivial $P$, the theorem gives a direct nonlinear analogue of the
X-ray transform.  It also gives the appropriate generalisation of the
scattering transform for ASDYM fields with arbitrary $P$ as then the
space $\cH_{G_\C}$ is the appropriate algebreo-geometric
generalisation of the solitonic data, and $H$ is the appropriate
generalisation of the reflection coefficient describing the
radiative/dispersive modes of the field.

The motivation for the study of scattering arises from a number of
areas.  It has long been suggested that scattering for integrable
systems in dimensions greater than 1+1 should be trivial partly as a
consequence of Huygens principle, and as a consequence of calculations
in perturbation theory.  This is very much not the case for Ward's
integrable chiral model in $2+1$ where there is right-angle scattering
of lumps and this is a symmetry reduction of the ASDYM equations in
split signature.  Scattering for the ASDYM equations is trivial
perturbatively in Lorentz signature, but in the complex, and in
particular in split signature, the relevant amplitudes do not
vanish. Indeed it follows from the theorem above that in the case of
trivial $P$ but non-trivial $H$, if one makes a choice of past and
future infinity, although the data on one of future or past infinity
determines the whole field, the field obtained at future infinity is
different from that at past infinity and this is here developed
into a study of the scattering problem.  The original motivation of
the author was to make some contact with the perturbative calculations
of scattering in Witten (2004) but the ASD sector is effectively
suppressed there (as appropriate for scattering in Lorentz signature).

The principal tool is a generalisation of the Ward construction
adapted to split signature.  The standard Ward
construction uses a correspondence between points in space-time and
Riemann spheres (complex projective lines) in twistor space and
encodes the original ASDYM field into a holomorphic
vector bundle over (a region in) $\PT$. This generalisation uses a
correspondence between points of space-time and certain holomorphic
discs in twistor space with boundary on the real slice $\PT_\R$.  

The paper is constructed as follows.  In \S\ref{background} the basic
geometry of compactified ultrahyperbolic Minkowski space $\M$ and its
double cover $\widetilde\M$ is set up together with its correspondence
with complex twistor space $\PT$ and its real slice
$\PT_\R$.  In particular points of $\widetilde\M$ correspond to
oriented lines in $\PT_\R=\RP^3$ that bound holomorphic discs in $\PT$.
In \S\ref{globalsolutions} the main result Theorem \ref{mainthm2} is
proved and some of its consequences are explored.
In \S\ref{examples} we discuss various examples corresponding to
different choices of the topology of $E$ etc., and those solutions
that correspond to the pullback of ASDYM fields that are pulled back
from $\M=S^2\times S^2/\Z_2$.  Only the abelian examples are given
explicitly in this section---the non-abelian examples are discussed
and shown to exist here, but are only given in full detail in an
appendix as a Kahler formalism is required to express the solutions
straighforwardly and this is only presented in the appendix.  

In \S\ref{scattering} the construction is applied to the task of
calculating the scattering of characteristic data from past null
infinity, $\scri^-$, to future null infinity, $\scri^+$ when the data
is small.\footnote{The division of null infinity $\scri$ into future
and past is not canonical in split signature, but we will see that the
scattering problem can nevertheless be made sense of.}  The scattering
can be expressed simply in terms of certain holonomies associated to
the connection that is presented as initial data.  The holonomies $h$
of the connection around a 3 parameter family of loops (two oriented
loops for each point in real twistor space) encodes the original data
when both are small by a theorem of Novikov (2002).  This $h$ is then
related to the twistor data as derived in \S\ref{globalsolutions} by a
Birkhoff factorization on a family of lines in the twistor space.
Generically, the scattering will be nontrivial in the non-abelian
case, being $h\rightarrow h^{-1}$.  In the abelian case it reduces
simply to a sign reversal of the connection, but will be non-trivial
in the non-abelian case, requiring a sequence of two Birkhoff
factorizations to calculate the effect on the potentials.  It is worth
noting that in perturbation theory, the amplitudes for the self-dual
sector formally vanish in Minkowski signature, but are non-trivial in
the complex and in particular in split signature.

In \S\ref{conclusions} some further avenues are discussed.  In the
first apendix a technical lemma required in the proof of theorem
\ref{mainthm} is proved.  In the second a formalism for the twistor
correspondence for $\widetilde\M$ is developed adapted to a choice
of a complex structure on $\widetilde\M$ given by an identification
with $\CP^1\times\CP^1$.  This is then used to give an explicit
description of the ADHM construction adapted to split signature to
give global solutions with second Chern class 2, and to work through
an example of the Ward ansatze.  

Finally we give some references to earlier and related work on these
issues.  Lerner (1992) introduced a similar such $H$ to describe ASD
Yang-Mills fields in split signature but did not fix the global
behaviour and so does not incorporate the bundle $E$ or the gauge
fixing that arises when $E$ is trivial.  The methods used here are a
development of those presented in Mason (1995) and \S10.5 of Mason \&
Woodhouse (1996) (which emphasized a non-Hausdorff twistor space
construction).  A key improvement in this paper is the use of
holomorphic discs with boundary on $\RP^3$ and this arises from
analogous work in lower dimension with Claude LeBrun on Zoll
projective structures in dimension 2, LeBrun and Mason (2002).
Similar methods apply to the split signature version of the nonlinear
graviton construction, Penrose (1976), for anti-self-dual conformal
structures on $S^2\times S^2$ and this is treated in a separate joint
paper with Claude LeBrun, LeBrun \& Mason (2005).  

\smallskip

\noindent 
{\bf Acknowledgements:} I would like to thank Claude LeBrun for a
number of important contributions to this work.  Thanks are also due
to Toby Bailey, Maciej Dunajski, Mike Eastwood, Gavin Kelly, Elmer
Rees, George Sparling, Nick Woodhouse and the anonymous referee.  I
would also like to thank the Department of Mathematical Sciences at
the University of Edinburgh for hospitality for some of the time while
this work was being written up.

\section{Twistors in split signature}\label{background}

\subsection{Conformally compactified Minkowski space}
We denote signature $(p,q)$ Minkowski space by $\R^{p,q}$ which is
$\R^{p+q}$ with a flat metric of signature $(p,q)$.  The standard
conformal  compactification $\M^{p,q}$ of $\R^{p,q}$ is obtained by
adding a 
`lightcone at infinity', denoted $\scri$ and has a standard
representation as the projectivisation of the lightcone of the origin
of $\R^{p+1,q+1}$. We will only be concerned with $\M^{2,2}$ which we
will denote by $\M$: it is a projective quadric of signature $(3,3)$
in $\RP^5$.  It is easily seen to have topology $S^2\times S^2/\Z_2$
by choosing coordinates $(\bx ,\by)$ on $\R^6$,
$\bx ,\by\in\R^3$, such that the quadratic form is
$Q=\bx \cdot\bx -\by\cdot\by$ and $\cdot$ denotes the standard
positive definite inner product on $\R^3$.  Then the light cone of the
origin is given by $Q=0$ which gives $
\bx \cdot\bx =\by\cdot\by$ and we can normalize
$\bx \cdot\bx =\by\cdot\by=1$.  Clearly this gives $S^2\times
S^2$ in $\R^6$, and the projection to $Q=0$ in $\RP^5$ is the quotient
by the joint antipodal map $\tilde\sigma$ on each $S^2$,
$\tilde\sigma:(\bx ,\by)\rightarrow(-\bx ,-\by)$.  The conformal
structure on $\M$ is determined by requiring that the lines in $\RP^5$
that lie on $Q=0$ are the null geodesics of $\M$.

We will also be interested in the double cover,
$\widetilde\M=S^2\times S^2$ of $\M$ with covering map
$\sigma:\widetilde\M\rightarrow\M$.  Concretely we can parametrize
$\widetilde\M$ by complex stereographic coordinates $w_1$ and $w_2\in
\C$ so that $$
\bx=\frac{(w_1+\bar w_1, i\bar w_1-iw_1,
1-|w_1|^2)}{(1+|w_1|^2)}\quad \mbox{ and }\quad \by=\frac{(w_2+\bar w_2, i\bar w_2-iw_2,
1-|w_2|^2)}{(1+|w_2|^2)}$$ 
and 
%spherical polars $(\theta_1,\phi_1,
%\theta_2,\phi_2)$ such that
%$$\bx=(\sin\theta_1\cos\phi_1,\sin\theta_1\sin\phi_1,\cos\theta_1)\, , \qquad
%\by=(\sin\theta_2\cos\phi_2,\sin\theta_2\sin\phi_2,\cos\theta_2),$$
%with metric 
\be{s2xs2} 
\d s^2=
\frac{4\d w_1\d\bar w_1}{(1+|w_1|^2)^2} - \frac{4\d w_2\d\bar
  w_2}{(1+|w_2|^2)^2} 
%\d\theta^2_1+\sin^2\theta_1\d\phi^2_1-\d\theta^2_2-\sin^2\theta_2\d\phi^2_2 
\, .  \ee If we take the volume form to be\footnote{In LeBrun \& Mason
2005 the opposite and more natural sign is taken for the volume form;
the conventions used here are consistent with those of Mason \&
Woodhouse (1996).}
$$
\d \mbox{Vol}=
\frac{\d w_1\wedge \d \bar w_1\wedge\d w_2\wedge \d \bar
w_2}{(1+|w_1|^2)^2(1+|w_1|^2)^2 }
%\d\theta_1\wedge\sin\theta_1\d\phi_1 \wedge
%\d\theta_2\wedge\sin\theta_2 \d\phi_2
$$
then the self-dual 2-forms are spanned by \be{sd2forms}
%\d(\theta_1\pm \theta_2)\wedge (\sin\theta_1\d\phi_1 \pm
%\sin\theta_2\d\phi_2)
\d w_1\wedge\d w_2\, , \quad\d \bar w_1\wedge\d \bar w_2\, ,
\quad \mbox{ and } \quad
\frac{4\d w_1\wedge\d\bar w_1}{(1+|w_1|^2)^2} - \frac{4\d w_2\wedge\d\bar
  w_2}{(1+|w_2|^2)^2} 
%\d\theta_1\wedge\d\theta_2+
%\sin\theta_1\sin\theta_2\d\phi_1\wedge\d\phi_2 
\ee and the
anti-self-dual forms by \be{asd2forms}
\d w_1\wedge\d \bar w_2\, , \quad\d \bar w_1\wedge\d w_2\, ,
\quad \mbox{ and } \quad
\frac{4\d w_1\wedge\d\bar w_1}{(1+|w_1|^2)^2} + \frac{4\d w_2\wedge\d\bar
  w_2}{(1+|w_2|^2)^2} \, .
%\d(\theta_1\pm \theta_2)\wedge
%(\sin\theta_1\d\phi_1 \mp \sin\theta_2\d\phi_2)\, , \quad \mbox{ and }
%\quad \d\theta_1\wedge\d\theta_2-
%\sin\theta_1\sin\theta_2\d\phi_1\wedge\d\phi_2.  
\ee 

\subsection{Twistors}
The twistor correspondence is the real Klein correspondence in which
each point $p\in\M$ corresponds to a line $L_p$ in the real twistor space
$\PT_\R=\RP^3$ (here $\T_\R$ denotes real non-projective twistor
space, $\R^4$ and $\T$ its complexification, $\C^4$).  The
  correspondence follows by representing a line in $\PT_\R$ by a
  2-plane through the origin in $\T_\R$ and then parametrizing such
  2-planes by simple bivectors $X\in\wedge^2\T_\R=\R^6$ up to scale.
  The simplicity condition is $X\wedge X=0$ which defines the
  quadric $\M\subset \P(\wedge^2\T_\R)=\RP^5$.
Under this correspondence, points of $\widetilde\M$ correspond to
oriented lines in $\PT_\R$.

Twistor theory makes essential use of the complexification $\PT=\CP^3$
of $\PT_\R$.  Each point $x\in\widetilde\M$ corresponds to a
holomorphic closed disc $D_x\subset\PT$ lying in the complexification $\C
L_{\sigma(x)}$ of $L_{\sigma(x)}$ (denoting the image of $x$ in $\M$
by $\sigma(x)$) such that $\p D_x=L_{\sigma(x)}$ and so that the
induced complex structure on $D_x$ induces the appropriate orientation
on the boundary corresponding to $x$.
  
A real twistor $Z\in\PT_\R$ (resp.\ dual twistor $W\in\PT_\R^*$)
corresponds in $\M$ to a totally null self-dual (resp.\ 
anti-self-dual) two-plane, referred to as an $\alpha$-plane (resp.\ 
$\beta$-plane) corresponding to the lines in $\PT_\R$ through $Z$
(resp.\ lines lying in the plane corresponding to $W$).\footnote{In
  $\widetilde \M$ the general $\alpha$-plane, with our conventions, is
  the
  graph of an orientation reversing isometry from one $S^2$ factor to
  the other, and the general $\beta$-plane is the graph of an
  orientation preserving isometry from one $S^2$ factor to the other.
    %Real twistor space, $\PT_\R=\RP^3$ is the space of such $\alpha$-planes.  
  The above representation identifies real twistor space $\PT_\R$ with
  $\SO(3)=\PSU(2)$.  Complex non-projective twistor space $\T$ can be
  represented as the space of complex non-vanishing $2\times 2$
  matrices with real slice $\T_\R$ being given by those that are
  unitary up to a real scale.  This description is taken further in
  the appendix.}

In order to expedite the correspondence, we introduce the
(six-dimensional) correspondence space $\F=\{(x,Z)\in
\widetilde\M\times\PT | Z\in D_x\}$ which naturally fibres over both
$\PT$ and $\widetilde\M$. 
\begin{equation}\label{doublefibration}
\begin{array}{rlcrl}
&&\F&&\cr
&p\swarrow&&\searrow q&\cr
\widetilde\M&&&&\PT
\end{array}
\end{equation}
The fibre of $p:\F\rightarrow \widetilde\M$
at $x\in\widetilde\M$ is the corresponding disc $D_x\subset\PT$.

The real correspondence space $\F_\R=\{(x,Z)\in
\widetilde\M\times\PT_\R | Z\in \p D_x\}$ is the 5-dimensional
boundary of $\F$ and fibres over $\PT_\R$ with fibres consisting of
lifts of $\alpha$-planes with topology $S^2$.  For the fibre of
$q:\F\rightarrow\PT$ at $Z\not\in \PT_\R$ we have
\begin{lemma}
 $q:\F-\F_\R\rightarrow\PT-\PT_\R$ is $1:1$ and
onto.  
\end{lemma}
\proof This follows from the fact that, given $Z\in\PT-\PT_\R$, there
is a unique real line $L\subset \PT_\R$ whose complexification $\C L$
contains $Z$ since $\C L$ must also contain $\bar Z$ and therefore be
the line joining $Z$ to $\bar Z$.
Furthermore, $Z$ determines the disc $D\subset\C L$ containing $Z$ with
boundary $L$ and so this $D$ corresponds to an $x\in\widetilde
\M$.  Thus $Z$ determines $x$ together with $Z\in D_x$.

\begin{corol} $\F-\F_\R$ has a natural complex structure from its
identification with $\PT-\PT_\R$.
\end{corol}

\rmk Analogous to the Atiyah-Hitchin-Singer definition of twistor
space, we note that $\F-\F_\R$ has a natural interpretation as the
bundle of metric compatible complex structures on $\widetilde\M$ and
its complex structure can be defined as in Atiyah, Hitchin and Singer
(1978).  However, on $\F_\R$ the distribution defining the complex
structure has a real part and so the description breaks down.

\subsection{Coordinates on an affine chart}\label{standardcoords}
If we we send the light cone $\scri$ of a point $i$ to $\infty$ we are
left with an affine chart $\R^{2,2}\subset\M$ on which we can
introduce standard Penrose notation, Penrose \& Rindler (1984 \& 1986)
adapted to split signature.  Here we take the corresponding points in
$\widetilde\M$ to be $i^-$ given by $(w_1,w_2)=(0,0)$ or its antipode
$i^+=(\infty,\infty)$ and $\scri$ is the hypersurface $|w_1|=|w_2|$.
Taking out $\scri$ divides $\widetilde\M$ into two copies
$\M^{\pm}=\{\pm(|w_1|-|w_2|)>0\}$ of $\R^4$, one of which, say $\M^+$,
can be taken to be `physical' space-time.  We use affine coordinates
on $\M^+$ which can be expressed in terms of
$(\bx,\by)=(x_1,x_2,x_3,y_1,y_2,y_3)$ as \be{affinecoords}
x^{AA'}=\frac{1}{\sqrt 2(x_3-y_3)}
\begin{pmatrix}x_1-y_1 &  x_2+y_2 \\ -x_2+y_2&x_1+y_1\end{pmatrix}\, ,
\qquad A=0,1, \; A'=0',1'\, .  \ee Here spinor indices $A,A'$ are
raised and lowered with the skew-symmetric spinors
$\varepsilon_{AB}=-\varepsilon_{BA}$ and $\varepsilon_{A'B'}$,
$\varepsilon_{01}=\varepsilon_{0'1'}=1$ and so transform under
$\SL(2,\R)$.  The conformal structure can be represented by the metric
$$
\d s^2=\d x^{AA'}\d x^{BB'}\varepsilon_{AB}\varepsilon_{A'B'} \, .
$$
We can then introduce complex homogenous coordinates
$(\omega^A,\pi_{A'})$ on $\PT$ (which, for real values restrict to
real homogeneous coordinates on $\PT_\R$).  Then the correspondence is
given by the incidence relation \be{twiscorr}
\omega^A=x^{AA'}\pi_{A'}\ee which can be read either as an equation
defining the $\alpha$-plane in $\M^+$ for fixed $(\omega^A,\pi_{A'})$
or as an equation defining a projective line in $\PT_\R$ for fixed
$x^{AA'}$.

The holomorphic discs in $\PT$ with boundary on $\PT_\R$ that
correspond to points of $\M^\pm$ can be parametrized by complex
homogeneous coordinates $\pi_{A'}$ subject to $\pm
i\pi_{A'}\bar\pi^{A'}\geq 0$ and $\omega^A$ given by (\ref{twiscorr}).
In the $+$ case, the homogeneous coordinates are related to the
standard disc coordinate $z$ with $|z|\leq 1$ by $z=(\pi_{0'} + i
\pi_{1'})/(\pi_{0'}-i\pi_{1'})$.

The spin bundle $\bbS$ has coordinates $(x^{AA'},\pi_{A'})$ and
the restriction of $\F$ to $\M^+$ can be identified with the subset of
the projective spin bundle $\P\bbS$ on which $i\pi_{A'}\bar\pi^{A'}\geq
0$ with equality on $\F_\R$.  The fact that the map from $\F-\F_\R$ is
1-1 can be expressed as the fact that given $(\omega^A,\pi_{A'})$ with
$i\pi_{A'}\bar\pi^{A'}>0$ equation (\ref{twiscorr}) has the unique real
solution
$$
x^{AA'}=\frac{\omega^A\bar\pi^{A'}+\bar\omega^A\pi^{A'}}{i\pi_{B'}
  \bar\pi^{B'}}\, . 
$$

On $\bbS$ we define the {\em twistor distribution}
$\cD=\{\p/\p\bar\pi_{A'} , \pi^{A'}\p_{AA'}\}$ and this descends also
to $\P\bbS$. This is conformally invariant.  On $\F-\F_\R$
$\dim\{\cD\cap\bar\cD\}=0$, the projection $q$ is $1:1$ and
and $\cD$ descends to give $T^{0,1}$ for the standard complex
structure on $\PT-\PT_\R$.  The d-bar operator can be written
\be{d-bar} \bar\p=\d\bar\pi_{A'}\frac{\p}{\p\bar\pi_{A'}} +
\frac1{\pi^{A'}\bar\pi_{A'}}\d x^{AA'}\bar\pi_{A'}\pi^{B'}\p_{AB'} \ee
On $\F_\R$, $\dim\{\cD\cap\bar\cD\}=2$ and
$\cD\cap\bar\cD=\{\pi^{A'}\p_{AA'}\}$ is then tangent to the
2-dimensional fibres of the projection $q:\F_\R\rightarrow\PT_\R$.

%If we consider functions that are holomorphic in $\pi_{A'}$, then the
%key part of the $\bar\p$-operator is $\pi^{A'}\p_{AA'}$.  

\subsection{The ASDYM equations and Lax pair}\label{Laxpair}
The ASDYM equations are equations on a connection $D$ on a bundle
$E'\rightarrow \widetilde \M$.  In a given local trivialisation
$D=\d+A$ where $A=A_{AA'}\d x^{AA'}\in\Omega^1(\widetilde\M)\otimes
\mathrm{u}(n)$ for a $\U(n)$ connection.  The curvature is $F=D^2=\d
A+ A\wedge A$.  The ASDYM equations are the condition that the
curvature satisfies the anti-self-duality condition $F^*=-F$ where $*$
is the Hodge dual, $F^*_{ab}=\half\varepsilon_{ab}{}^{cd}F_{cd}$.  The
curvature naturally decomposes into its self-dual and anti-self-dual
parts when expressed in spinors
$$
F_{AA'BB'}=\varepsilon_{AB}\phi_{A'B'}+\varepsilon_{A'B'}\phi_{AB}\, ,
$$ where $\phi_{A'B'}=\p_{(A'}^A A_{B')A}+A_{(A'}^A A_{B')A} $
(resp. $\phi_{AB}\p_{(A}^{A'} A_{B)A'}+A_{(A}^{A'} A_{B)A'}$) is the
self-dual (resp.\ anti-self-dual) part of the curvature.
A connection is ASD iff $\phi_{A'B'}=0$.

A Lax pair for the ASDYM equations on $D$ is given by $\pi^{A'}D_{AA'}$ in
the sense that $[\pi^{A'}D_{AA'},\pi^{B'}D_{BB'}]=0$ iff the ASDYM
equations hold  (Ward 1977).

\section{Global ASDYM fields in
  split signature}\label{globalsolutions} 
\subsection{The generalised Ward correspondence}
Our main theorem for $\U(n)$ ASDYM fields on
$\widetilde\M$ is as follows
\begin{thm}\label{mainthm}
  There is a 1-1 correspondence between smooth $\U(n)$ ASDYM fields on
  $\widetilde\M$ and pairs $(E,H)$ where $E$ is a rank $n$ holomorphic
  vector bundles on twistor space $\PT$ and $H$ is a smooth positive
  definite hermitian metric on the fibres of $E|_{\RP^3}$.

  If there exists an antilinear conjugation
  $\tilde \sigma_E:E\rightarrow \bar E^*$ covering the standard complex
  conjugation $\tilde\sigma :\PT\rightarrow\PT$ that fixes $\PT_\R$,
  such that $H$ is induced by $\tilde \sigma_E$ by $H(v,v)=
  (\tilde\sigma_E v)(\bar v)$ then
  the ASDYM field on $\widetilde\M$ is one that is pulled back from
  $\M$.
\end{thm}
A technical lemma required in the proof of this theorem is relegated to
the first appendix.

\smallskip

\noindent
{\bf Proof (forward direction):} We are given an ASDYM connection
$D=\d +A$ on the bundle $E'\rightarrow \widetilde\M$ where $ A $ is a
1-form on $\M$ with values in the Lie algebra of $\U(n)$.  In order to
produce the pair $(E,H)$ where $E$ is a holomorphic vector bundle over
$\PT$ and $H$ is a Hermitian metric on $E|{\PT_\R}$, define first
$E\rightarrow \PT-\PT_\R$ by $E= (q^{-1})^*p^*E'$ where $p$ and $q$
are the projections of the double fibration (\ref{doublefibration}).
The connection $D$ lifts to give a connection on $p^*E'\rightarrow
\F$, and, away from $\p\F$, the map $q$ is $1:1$ and hence this
determines a connection on $E= (q^{-1})^*p^*E'$ over $\PT-\PT_\R$.
The connection determines a $\bar\p$-operator on $E$ which can be
represented in the affine coordinates above as
$$
\bar\p_E=\d\bar\pi_{A'}\frac{\p}{\p\bar\pi_{A'}} +
\frac1{\pi^{A'}\bar\pi_{A'}}\d x^{AA'}\bar\pi_{A'}\pi^{B'}D_{AB'}\, .
$$
It is a standard calculation that $\bar\p_E^2= 0$ as a consequence
of the ASDYM equations based on 
\S\ref{Laxpair} and so $(E,\bar\p_E)$ is a holomorphic vector bundle
over $E|_{\PT-\PT_\R}$.

We now define $E\rightarrow \PT_\R$ to be the bundle whose fibre at
$Z\in\PT_\R$ is the space of covariantly constant sections over the
corresponding $\alpha$-plane in $\widetilde\M$ (the $\alpha$-planes in
$\widetilde\M$ are simply connected, being $S^2$s, and the
anti-self-duality condition implies that the curvature vanishes on
each real $\alpha$-plane so the space of covariantly constant sections
is well defined).  Clearly, $E|_{\PT_\R}$ carries a hermitian metric
$H$.

The definitions of $E$ over $\PT-\PT_R$ and over $\PT_\R$ are
quite different but we have
\begin{lemma}\label{techlemma} 
The given definition of $E\rightarrow
\PT_\R$ is a smooth extension of $E\rightarrow\PT-\PT_\R$ such that the
d-bar operator $\bar\p_E$ extends smoothly over $\PT_\R$.
\end{lemma}
The proof of this  is
relegated to an appendix.
With this, $E$ extends over all of $\PT$ and its restriction to
$\PT_\R$ has a naturally defined 
Hermitian metric $H$. 

\smallskip

\noindent
{\bf Proof (backward direction):} Starting with a pair $(E,H)$, we
wish to construct an anti-self-dual Yang-Mills field on $\widetilde
\M$.  We first construct a principal $\U(n)$ bundle
$P'\rightarrow\widetilde\M$ whose fibre $P'_x$ at $x\in\widetilde \M$ is
$$
P'_x=\{\mbox{holomorphic frames $g$ of }E\rightarrow D_x\, |\, 
g\mbox{ is unitary w.r.t. $H$ on $\p D_x$}\}
$$ In order to see that this is well defined, we first choose a Stein
neighbourhood $U$ of $D_x$ in $\PT$ and choose a holomorphic
trivialization of $E$ over $U$.  By an abuse of notation, denote by
$g$ and $H$ the matrices representing the frame $g$ and Hermitian
metric $H$ in this trivialisation of $E$.  Then $g(x,Z)$ must be
holomorphic on $D_x$ and satisfy
\be{mainfact}
gHg^*=1
\ee 
on $\p D_x$.  This is a Birkhoff factorization of $H$ as $g^*$
extends naturally to a holomorphic function on $\C L_x -D_x$ since
complex conjugation on $\PT$ restricts to $\C L_x$ sending $D_x$ to
$\C L_x -D_x$.  In order for $g$ to be well defined we first need to
know that this Birkhoff factorization always has trivial homomorphism
factor from $\p D_x\rightarrow \GL(n)$ (i.e., the matrix $H|_{\p D_x}$
to be factorized always lies in the `big cell').  This follows from
the fact that $H$ is positive definite, see Gohberg \& Krein (1958) or
Mason \& Woodhouse (1996) proposition 9.3.6.  Thus, a Birkhoff
factorization $gH\tilde g=1$ exists for some $g, \tilde g$ holomorphic
on $D_x$ and $\C L_x- D_x$ respectively, unique up to $g, \tilde g
\rightarrow Ag, \tilde g A^{-1}$ for some constant matrix $A$ .  We
need to show that we can choose $A$ so that $\tilde g =g^*$.  We first
note that since $H$ is Hermitian, we have that $\tilde g^* H g^*=1$.
Therefore, eliminating $H$, on $\p D_x$ we have $g\tilde g^{*-1}=
\tilde g^{-1} g^*$.  However, the left hand side of this equation can
be continued holomorphically over $D_x$, whereas the right continues
holomorphically over $\C L_x -D_x$, so together they define a matrix
valued function that is global on the Riemann sphere, and hence
constant by Liouville's theorem. Under $g,
\tilde g \rightarrow Ag, \tilde g A^{-1}$ we have $g\tilde g^{*-1}
\rightarrow Ag\tilde g^{*-1}A^*$ and, since $g\tilde g^{*-1}$ is
Hermitian,  $A$ can be chosen to reduce $g\tilde g^{*-1}$ to the
identity with the residual freedom of $A$ such that $AA^*=1$, i.e.,
the unitary group.  Therefore, $P'\rightarrow \widetilde\M$ is well defined and
naturally has the structure of a principal $U(n)$ bundle.

We now wish to construct an ASDYM connection on $P'$. By construction, 
there are natural trivialisations of $P'$ over the $\alpha$-planes
in $\widetilde \M$ obtained by choosing, for $Z\in\PT_\R$, a unitary
frame $g_Z$ of $E_Z$, and requiring that, for each $x$ with $Z\in \p
D_x$, $g(x,Z)=g_Z$.  
\begin{lemma}
There exists a unique connection on $P$ up to gauge transformations
such that the above frames are be covariantly constant.  Such a
connection is necessarily 
anti-self-dual.
\end{lemma}
We construct the connection on $P$ using $g$ as follows.  The
expression $g^{-1}\pi^{A'}\nabla_{AA'}g$ is holomorphically defined as
a section of $\Oc(1)$ over each $D_x$ and is skew-hermitian on $\p D_x$
since $\pi^{A'}\p_{AA'}$ is real and $g$ is unitary there.  (Here
$\Oc(1)$ is the dual of the tautological line bundle $\bbS\rightarrow\PS$
over the projective spin bundle; on each $\CP^1$ fibre of $\PS$, it
restricts to the standard line bundle of Chern class 1 whose sections
can be represented by homogeneous functions of degree 1). It is
therefore equal to $A_{AA'}\pi^{A'}$ for some skew-Hermitian $A_{AA'}$
depending only on $x\in U'$ by an extension of Liouville's theorem.
The extension of Liouville's theorem in question states that a
holomorphic function on the unit disc that is real on the boundary is
necessarily a real constant.  This follows from the standard Liouville
theorem by extending the function over the whole complex plane by
setting $f(z)=\overline{f(1/\bar z)}$ and noting that the resulting
function is continuous and hence holomorphic on $|z|=1$ since it is
real there and so is a bounded holomorphic function on $\C$.  This
extends to sections of $\Oc(1)$ by considering
$(\pi_{0'})^{-1}g^{-1}\pi^{A'}\nabla_{AA'}g$ which has a simple pole
at $\pi_{0'}=0$, but is otherwise, via the above argument, holomorphic
on the Riemann sphere and is therefore equal to a skew Hermitian
$(\pi_{0'})^{-1}(A_{A0'}\pi_{1'} -A_{A1'}\pi_{0'})$ where $A_{AA'}$
depends only on $x$.$\Box$

In the case that $H$ is induced by an anti-holomorphic map
$\tilde\sigma_E:E\rightarrow \bar E^*$ covering the standard
complex-conjugation, the ASDYM field can be constructed directly on
$\M$ via the standard Ward transform and the fact that it will give
rise to a real ASDYM field follows from the reality structure
$\tilde\sigma_E$. The construction works by defining, for $x\in\M$,
$E'_x= \Gamma (\C L_x,E)$ with Hermitian form induced by
$\tilde\sigma_E$.  The construction of the connection follows as above
replacing $g$ by the expression for a frame of $E'_x= \Gamma (\C
L_x,E)$ that is unitary on the real slice in some local trivialisation
of $E\rightarrow \PT$ on a neighbourhood of $D_x$. $\Box$

\medskip\noindent {\bf Remarks.} \\ 1. This theorem can easily be
extended to any compact gauge group $G$ by embedding $G$ in $\U(n)$
for some $n$.  The construction is most easily stated in terms
of principal bundles as in the introduction:
\begin{thm} Gauge equivalence classes of $C^\infty$ solutions to the
  ASDYM equations with gauge group $G$ are in 1-1 correspondence
  between principal $G_\C$ bundles $P\rightarrow \PT$ together with a
  $C^\infty$ reduction of the structure group to $G$ over $\PT_\R$
  where $G_\C$ is the complexification of $G$. The reduction of the
  structure group over $\PT_\R$ can be alternatively expressed as a
  $C^\infty$ section $H:\PT_\R\rightarrow P/G$ of the $G_\C/G$ bundle
  $P/G$.
\end{thm}

\noindent
2. Unlike the case of Euclidean signature, solutions to equations in
indefinite signature can have pretty much arbitrarily low regularity.
It is clear that, taking the correspondence in the reverse direction,
the bundle $P$ or $E$ over $\PT$ is always necessarily analytic by
elliptic regularity, but $H$ can be chosen to have any regularity for
which the Birkhoff factorization will work.  If we work in a H\"older
space framework, the regularity of solutions to the Birkhoff
factorization problem have regularity $C^{k,\alpha'}$ when $H$ has
regularity $C^{k,\alpha}$ for $\alpha'<\alpha$ assuming that there are
no jumping lines (as is the case here).  This follows from the
corresponding results for the Hilbert transform and the implicit
function theorem.  The connection on space-time is therefore of
regularity $C^{k-1,\alpha'}$ since one further derivative is taken.
This suggests that given an ASDYM connection on space-time of given
regularity, $H$ can be shown to have one extra degree of regularity.
However, the construction of $H$ from teh connection in the proof
loses many degrees of regularity so these methods will not yield a
definitive theorem.

\subsection{The topology of $E$}  The theorem 
implies that {\em any} holomorphic vector bundle $E\rightarrow \PT$ with
hermitian metric $H$ on $E|_{\PT_\R}$ will give rise to anti-self-dual
Yang-Mills fields with gauge group $\U(n)$ so long as the hermitian
metric $H$ is positive definite---we do not need to concern ourselves
with `jumping lines'.

The topology of holomorphic vector bundles on $\CP^3$ are
characterised by the Chern classes $c_1$, $c_2$ and $c_3$ except in
the case of rank 2 bundles for which $c_3$ is trivial, but when $c_1$
is even, can admit a mod 2 `$\alpha$-invariant,' Atiyah \&
Rees (1976).  All such invariants can be non-trivial in
contradistinction with the case of instantons on $S^4$, for which the
only possible non-trivial topological invariant is the second Chern
class of the original Yang-Mills vector bundle on $S^4$ which is the
same as the second Chern class of $E\rightarrow\CP^3$.  Here $c_1$ and
$c_3$ can be non-trivial also.

Firstly $c_2(E')(\widetilde\M)=2c_2(E)(\CP^2)$ since
$c_2(E')(\widetilde\M)$ can be represented as the integral of
$(-1/8\pi^2)\tr F^2$ over $\widetilde\M$.  This integral is the same
as that over the quadric
$\sum_\alpha (Z^\alpha)^2=0$  in $\PT$ since this is a section of the
fibration $\F-\p\F=\PT-\PT_\R\rightarrow\widetilde\M$ (this geometry
is explained more fully in appendix 2).  However, the quadric
is twice the generator of the 2nd cohomology of $\PT$.

Furthermore non-trivial $c_1(E)$ is also allowed.  This arises from
$c_1(E')$ since $c_1(E)[\mbox{line}]=c_1(E')[\beta\mbox{-plane}]/2$
using the identification in the previous paragraph since the
$\beta$-plane becomes identified with a conic obtained by intersecting
a plane in $\PT$ with the quadric.  Note that since $E'$ is flat on
$\alpha$-planes, $c_1(E')[\alpha\mbox{-plane}]=0$ so the possible
first Chern classes are measured just by evaluation on $\beta$-planes
and must be even from above. $\U(1)$ examples with nontrivial $c_1$
will be given in the next section.

More remarkably $c_3(E)$ can be non-trivial.  For general gauge group
there is a mod 2 relation on $c_3$, i.e., if $c_1=c_2=0$, $c_3$ must
be even (Rees private communication), although for $\SU(2)$ there is
only the $\alpha$-invariant.  At least in the $\SU(2)$ case
it is known that this invariant can be non-trivially realised with a
holomorphic vector bundle for any given even $c_1$ and arbitrary
$c_2$, for example it is non-trivial for $E=\Oc(2)\oplus\Oc(-2)$
(although, of course, $c_2$ is also non-trivial for this bundle),
Atiyah \& Rees (1976).

Non-trivial  $c_3$ cannot  arise from  the topology  of $E'\rightarrow
\widetilde\M$  as   $\widetilde  \M$  is   4-dimensional  and  indeed,
non-trivial  third Chern  classes can  arise when  $E'$ is  trivial by
choosing  $c_1(E)=c_2(E)=0$  but $c_3(E)\neq  0$.   To  see where  the
topological non-triviality comes  from in this case, we  note that for
unitary groups, $E|_{\PT_\R}$ will be trivial so that a trivialization
of $E|_{\PT_\R}$  can be  pulled back to  $\F_\R$ and compared  to the
pullback of  a trivialisation $E'$.  The  gauge transformation between
these trivializations gives a map $g: \F_\R\rightarrow \U(n)$ and this
will be  topologically non-trivial when  $c_3(E)$ or $\alpha$  are non
trivial.\footnote{It was erroneously assumed that these classes should
vanish in Mason (1995).}

\subsection{The case when $E$ is trivial} \label{Etrivial}
For a `small' ASDYM field, the topological invariants of $E$ will
necessarily be trivial and this implies that $E$ is analytically
trivial.  If $E$ is trivial as a holomorphic vector bundle, a
holomorphic trivialization of $E$ is unique up to a global constant
$\GL(n,\C)$ transformation.  In such a global trivialisation,
$H:\RP^3\rightarrow \mbox{Herm}^+_n$ where $\mbox{Herm}^+_n$ denotes
the space of $n\times n$ positive definite Hermitian matrices and $H$
is defined up to $H\rightarrow gHg^*$ for constant $g\in\GL(n,\C)$.
Such equivalence classes of $H$ completely characterise ASDYM fields
on $S^2\times S^2$ with `small' data.  The correspondence between $H$
and the corresponding ASDYM field is a nonlinear analogue of the X-ray
transform (see below).

\section{Examples}\label{examples}
\subsection{The abelian case}\label{abelian}
Consider first the case where $E$ is the trivial line bundle.  Then
$H$ is simply a real non-vanishing function on $\PT_\R$ and the
problem of constructing the corresponding field on space-time proceeds
by means of standard twistor integral formulae with twistor function
$\log H$ that go back to Ward and Sparling, see Ward (1977).  In the
affine coordinates given before, the integral formula leads to the
following standard formula for the ASD Maxwell field
$$
\phi_{AB}(x)=\oint_{\omega^A=x^{AA'}\pi_{A'}} \frac{\p^2\log
  H}{\p\omega^A\p\omega^B} \;\pi_{C'}\d \pi^{C'}\, .
$$

It is worth noting that because $H^1(\PT,\Oc)$ is trivial, there are no
ASD Maxwell fields on $\M$.  The Maxwell fields obtained from the
construction above are odd under the antipodal map on $\widetilde \M$
since the integral above requires an orientation on the line $L_x$.

A further novelty is that $H$ is a smooth function on $\PT_\R$ unique
up to a constant rather than an element of a cohomology class which
would be the norm for a Penrose transform.  The transform here yields
a helicity shifted variant of the standard X-ray transform.

Naively, we can connect the $f=\log H$ with cohomology following
Atiyah (1979).  Take $f$ to be analytic and extend it to some
neighbourhood $U$ of $\PT_\R$.  On a neighbourhood $V$ of a line that
is divided into two parts by $U$, say $V=V_0\cup V_1$ with $V_0\cap
V_1=U\cap V$, it can be taken to be the Cech representative for a
cohomology class relative to that covering of $V$.  (Alternatively a
Dolbeault representative can be obtained by extending $f$ off $U$ as a
smooth but non-holomorphic function on $\PT$ and, on a neighbourhood
$V$ of a line, we can consider the dolbeault form $\alpha$ such that
$\alpha=0$ on $V_0$ and $\alpha=\bar\p f $ on $V_1$.)  Clearly such
descriptions fail globally since, as the space-time point does a
circuit around a null geodesic in $\M$, if we follow this path with
different choices of $V$, $V_0$ and $V_1$ will be interchanged and 
so $f$ would have to be identified with $-f$.  This construction also
fails to explain how the cohomological gauge freedom is fixed.

One way to understand the construction globally and cohomologically is
as follows.  First, if such a function $f=\log H$ defined up to a
constant is taken to be analytic, it can be thought of as a relative
cohomology element in $H^1_{\PT-U}(\PT)$ where $U$ is some small open
neighbourhood of $\PT_\R$ via the connecting homomorphism of the long
exact relative cohomology sequence
$$
\ldots   \rightarrow
H^1(\PT) 
\rightarrow H^1_{\PT-U}(\PT)\stackrel{\delta}{\rightarrow} H^0(U) \rightarrow
H^0(\PT) \rightarrow \ldots
$$
using the observation that $H^1(\PT) =0$ and $H^0(\PT)=$ constants.
The relative cohomology then has a natural pairing with each $(D_x,\p
D_x)$ which is realised by the usual integral formulae. This ties in
with a point of view developed in Mason (1995) in which the twistor
space is taken to be the non-Hausdorff space obtained by gluing
together two copies of $\PT$ along $U$.  Cohomology on the space is
given by the relative cohomology group $H^1_{\PT-U}(\PT)$ by a result
of Bailey (1985), see also \S6 of Mason \& Hughston (1990).

The detailed correspondence between the Penrose transform and the
X-ray transform has been much studied elsewhere, Woodhouse (1992),
Mason (1995), Sparling (1998), Bailey, Eastwood, Gover and Mason
(1999, 2003), Bailey \& Eastwood (2001).  This connection between the
Penrose transform and X-ray transform is also used 
in twistor string theory, Witten (2004) and Berkovits and
Witten (2004).

\medskip
\noindent
{\bf The case where $E$ is nontrivial:} If $E$ is a nontrivial line
bundle, it must be isomorphic to $\Oc(k)$ for some $k$.  We must then
take $H$ to be a (real) non-vanishing section of $\Oc(-2k)$ over
$\PT_R$. Taking the simplest case, $k=-1$, we can set $H=\sum_\alpha
(Z^\alpha)^2$.  This can be transformed explicitly to yield the
$\U(1)$ gauge field with curvature
$$
F=
\frac{2\d w_1\wedge\d\bar w_1}{\pi(1+|w_1|^2)^2} + \frac{2\d w_2\wedge\d\bar
  w_2}{\pi(1+|w_2|^2)^2} \, .
%i(\d\theta_1\wedge
%\sin\theta_1\d\phi_1 - \d\theta_2\wedge \sin\theta_2\d\phi_2).
$$
This gives rise to a gauge field with non-trivial
first Chern class on $\widetilde \M$.\footnote{This is an analytic
  continuation of the ASD Coulomb/Dirac monopole solution centred on the
  complex curves either $\bx=0= \by\cdot\by$ or
  $\by=0=\bx\cdot\bx$---the points of these curves correspond to the
  complex lines in $\PT$ that generate the given quadric, see \S I.6.2
  of Hughston \& Mason (1990).}  Clearly the metric $\e^fH^{-k}$ on
$\Oc(k)$ will give the sum of $-k$ times the above solution with that
described above using the standard twistor integral formula for $f$.

\subsection{The t'Hooft and Ward ansatze}\label{ward}
Examples of non-abelian ASDYM fields with gauge group $\SL(2,\R)$ on
$\widetilde\M$ are constructed from the t'Hooft ansatze, with
$E=\Oc(1)\oplus\Oc(-1)$ in \S10.5.2 of Mason \& Woodhouse (1996).

It is not so easy to encode the reality conditions for real $\SU(2)$
solutions with the t'Hooft ansatz (see, for example, the next
section).  Imposing a symmetry on $\widetilde \M$ along $\mathrm{Im}\,
w_1\p/\p w_1$
reduces the ASDYM equations to a Yang-Mills-Higgs system on 2+1 de
Sitter space that has been studied by Kotecha \& Ward (2001).  The
specific solution actually considered by Kotecha \& Ward is not smooth
when pulled back to $\widetilde\M$ (thus there is also a singularity
at infinity in 2+1 de Sitter space), but the ansatz can be adapted to
give smooth solutions on $\widetilde\M$.

Reformulating \S 7 of Kotecha \& Ward (2001)
but without imposing the symmetry, we consider the case where
$E=\Oc(k)\oplus\Oc(-k)$ and
$$
H= \begin{pmatrix} 2Q^{-1} \cosh f & \e^{- f} \\  \e^{-
    f} & Q\e^{-f} \end{pmatrix}
$$
where $Q\in H^0(\PT,\Oc(2k))$ is a polynomial of homogeneity $2k$ that
does not vanish on $\PT_\R$ and $f$ is a smooth function on $\PT_\R$.
A straightforward choice of $Q$ is $Q=(\sum_\alpha (Z^\alpha)^2)^k$
This can be used to give a solution by means of the Ward ansatze as
described in \S8.2.4 of Ward \& Wells (1990):  we note that 
$$
H=FR\, , \quad \mbox { where }  \quad F=
\begin{pmatrix} \e^f & 2Q^{-1} \cosh f \\0&  \e^{- f} \end{pmatrix}
 \quad \mbox{ and } \quad 
R= \begin{pmatrix} 0 & -1 \\ 1 & Q \end{pmatrix}
$$
and that $R:\Oc(k)\oplus\Oc(-k) \rightarrow \Oc(-k)\oplus \Oc(k)$
is a global map of vector bundles on $\PT$. So, as far as the Birkhoff
factorization is concerned, we are reduced to an example of the Ward
ansatze as detailed in \S8 of Ward and Wells (1990) for which the
reconstruction of the space-time ASDYM field can be implemented by
quadratures.\footnote{The Kotecha Ward solution has $k=1$, $Q$ given
  as above, and $f=\log Q/((Z^2)^2+(Z^3)^2)$, but $f$ is not smooth on
  all of $\PT_\R$ and so the pullback of the corresponding solution to
  $\widetilde\M$ is not smooth.}  The full calculations are performed
in appendix \S\ref{kwsoln} where a more detailed formalism is
established that expedites the calculations.

%A simple choice is for $k=1$, $Q$ as above and $f= \log Q/P$
%where $P$ is another positive definite quadratic form on $\T_\R$.
%In this case 
%$$
%F=\begin{pmatrix} P/Q & P/Q^2 + 1/P \\0&  Q/P \end{pmatrix}
%$$

\subsection{Split signature instantons}\label{2inst}
The case in which $H$ is induced from a reality structure on $E$ is in
effect the case where the solution on $\widetilde \M$ is pulled back
from one on $\M$ that is constructed from a global holomorphic vector
bundle over $\PT$ using the standard Ward construction.  This is what
we will mean by a `split signature instanton'---there is no analogue of
the Bogomolny bound for the action by the second Chern class in split
signature, and so the concept of an instanton is not well defined, but
these solutions are the ones that are defined purely algebraically
geometrically.  

This case reduces to the standard Ward construction in which, in order
to obtain a regular solution on $\M$, $E$ must be trivial on
restriction to the complex lines that are complexifications of real
lines.  This non-singularity is in any case a consequence of the
assumption that the reality structure on $E$ induces a
positive-definite hermitian structure on the restriction of $E$ to the
real slice since the triviality of $E$ over $L_x$ follows by the same
argument as given in the reconstruction of the ASDYM field from the
bundle.  This in particular implies that $c_1(E)=0$.  Furthermore $E$
must be semi-stable since if there is a destabilizing subsheaf, it
will agree with $\Oc(k)$, $k>0$, on a generic line and this will lead
to $E$ being non-trivial on a generic real line.  It is easy to see
that if it is semistable but not stable, the subsheaf generically
isomorphic to $\Oc(0)$ will give rise to a line subbundle of $E'$ on
space-time on which the connection is trivial and so the connection
will be reducible.  Thus, assuming the ASDYM connection on $E'$ is not
reducible, we may assume $E$ to be stable and the ADHM machinery may
be invoked to construct $E$ and the solution on space-time as
described in Atiyah (1979).

\begin{thm}
  Split signature instantons necessarily have even $c_2(E)$ and exist for
  $c_2(E)=2$.
\end{thm}
\proof 
In complexified Minkowski space, $\CM$ there is a hypersurface
$\Sigma$ on which instantons are singular (points of $\Sigma$
correspond to lines in $\CP^3$ on which $E$ fails to be trivial) and
we must choose our bundle $E$ so that, not only are the fields real,
but also this hypersurface has no real points.  The singular
hypersurface has degree $c_2(E)$ in $\P^5$ and determines $E$ at least
for $c_2(E)=1$ or $2$, Hartshorne (1978).  In the case $c_2(E)=1$, the
hypersurface is linear, and must be real for a real solution.  It is
easy to see that it must then intersect $\M$ and indeed this will be
the case for any odd $c_2(E)$ as an algebraic hypersurface of odd
degree will always have real points.

For $c_2(E)=2$, the singular hypersurface $\Sigma$ is a quadratic cone
on $\CP^5$ of rank 3 subject to a certain `Poncelet' condition,
Hartshorne (1978). For reality, the 3-plane in the kernel of this
quadratic form is the complexification of a real $\R^3\subset \R^6$
and the quadric defining $\M$ restricts to be either Lorentzian or of
definite sign on this plane.  The former cannot lead to smooth
solutions, so we focus on the latter in the following.  In terms of
our homogenous coordinates $(\bx,\by)\in\R^3\times\R^3$ on $\RP^5$ we
can, after an $\SO(3,3)$ transformation, express $\Sigma$ as
$S(\bx,\bx)=\bx\cdot\bx$ for some symmetric trace-free matrix $S$.
The `Poncelet condition' of the conic $S(\bx,\bx)=\bx\cdot\bx$ with
respect to the conic $\bx\cdot\bx=0$ requires that for an arbitrary
point $\bx_0$ on $\bx\cdot\bx=0$, if we construct the tangent $l_0$ at
$\bx_0$ and its intersections with $S(\bx,\bx)=\bx\cdot\bx$ at points
${\bf a_1}$ and ${\bf a_2}$, then the other tangents to
$\bx\cdot\bx=0$ through ${\bf a_1}$ and${\bf a_2}$ must meet on the
conic $S(\bx,\bx)=\bx\cdot\bx$.  This Poncelet condition is satisfied
iff the $\tr(S^2)=3/2$.  It is clear that this can be satisfied in
such a way that $\Sigma$ has no real points, i.e., with the
eigenvalues of $S$ less than 1.$\Box$

\smallskip

Although these solutions can all be constructed explicitly by means of
the t'Hooft ansatze based on a solution to the ultrahyperbolic wave
equation $\phi=\sum_{i=1}^3 \lambda_i/(x-x_i) ^2$, the $x_i$ must be
three points of the conic $\bx\cdot\bx=0$ tangent to the sides of a
Poncelet triangle as described above, and so cannot be real. Therefore
$\phi$ cannot be real and it will be difficult to identify the values
of $x_i$ and $\lambda_i$ that give rise to a real solution or
represent it in a real (unitary) gauge.

Instead, the ADHM construction is worked through in an appendix to
give an explicit formula for the connection (it is not presented here
as it requires a description of the twistor correspondence that is
only developed earlier in the appendix).

\section{Nonlinear scattering theory} \label{scattering}

In this section we construct the map from initial data at $-\infty$ to
final data at $+\infty$ at least in the context of small data giving
rise to small solutions (so that the topology of both $E$ and $E'$ can
be taken to be trivial).  The key idea rests on two facts. Firstly the
intersections of $\alpha$-planes in space-time with null infinity are
circles, and the flatness condition on $\alpha$--planes implies that
the holonomy around these circles must be trivial. Secondly, we find
that the characteristic data on past null infinity can be encoded into
holonomies in effect around two halves of such circles, one associated
to the part of the $\alpha$--plane in past null infinity and the other
in a fixed family of $\alpha$-planes associated to the `$t=0$'
hypersurface dividing past from future null infinity.  The scattering
map must then take these holonomies to their inverse on the
corresponding intersection of the $\alpha$--plane with future null
infinity. We need to prove however that these holonomies uniquely
determine and are determined by the initial data, the final data and a
compatible ASDYM field on $\M$, at least for small data.

The construction will involve taking the family of holonomies of the
connection and performing a sequence of two Birkhoff factorizations on
the holonomies, the first to find $H$ (which determines the solution
on all of $\widetilde\M$) and the second to find the final data.  The
sense in which the data is small is that for the corresponding twistor
data, the holomorphic vector bundle $E$ is trivial so that the
solution is encoded into $H$.  It is also small in the sense that
these Birkhoff factorizations do not jump.  This can be expressed as
an analytic smallness condition on the asymptotic characteristic data,
see Novikov (2002) for an explicit statement of the smallness condition.
These calculations are nevertheless fully nonlinear.

We will see that the the calculation can be performed on restriction
to certain 2-planes, $\beta$-planes, on the initial and final data
surface $\scri$, so there is no interaction between data posed on one
$\beta$-plane and another.

\subsection{The geometry}
In split signature, the global structure does not provide a canonical
decomposition into past and future since there are two time-like
directions and we must make an (unnatural) choice of one of them to
proceed.  This is possible, and we will see that we can define
scattering in this way in spite of the signature.

The choice of infinity in this conformally invariant context is also
arbitrary and we do this first. We choose the point
$i^-\in\widetilde\M$ at past infinity and denote its antipode by
$i^+$, future infinity.  The data at infinity will be posed on the
lightcone $\scri$ of $i^-$ which reconverges on $i^+$.  In order to
make a choice of past and future, we choose a linear hyperplane
$\Sigma$ in $\RP^5$ that is not tangent to $Q=0$, and separates $i^+$
from $i^-$ cutting $\scri$ into $\scri^\pm$.

In the coordinatisation $(w_1,w_2)$ of $\widetilde\M$ of equation
(\ref{s2xs2}), the points $i^-$ and $i^+$ can be taken to be
$w_1=w_2=0$ and $w_1=w_2=\infty$ respectively.  The light cone $\scri$
of $i^-$ is the hypersurface $|w_1|=|w_2|$ and reconverges at $i^+$.
We can coordinatise $\scri$ with $(w_1,\eta)\in \C\times S^1$ by setting 
$$
(w_1,w_2)=(w_1, \e^{i\eta}w_1)\, .
$$ $\scri$ divides $\widetilde\M=S^2\times S^2$ into two copies
$\M^\pm$ of affine $\R^4$ and we take $\M^+=\{(w_1,w_2); |w_1| >
|w_2|\}$ as physical space-time and discard $\M^-$.  A convenient
choice for $\Sigma$ is the hyperplane $x_3=0$ or equivalently
$|w_1|=1$ (i.e., $x_3/(x_3-y_3)$ is taken to be the time variable).
$\Sigma$ divides $\scri$ into $\scri^\pm=\{\pm(|w_1|-1)\geq 0$\} with
$\scri^-$ being past null infinity and $\scri^+$ future null infinity.
The antipodal map on $\widetilde \M$ sends $\scri$ to itself giving a
canonical identification between $\scri^+$ and $\scri^-$ (the
light-cone of a point of $\scri^-$ reconverges on the `antipodal'
point on $\scri^+$) and so we will be able to compare initial data on
$\scri^-$ to `final data' on $\scri^+$.

\subsection{The characteristic data}
Characteristic data for the ASDYM equations is a connection $ A $ on a
bundle $E'$ over $\scri^-$ that is flat on the $\alpha$-planes on
$\scri^-$.\footnote{On $\scri$ the two-surfaces on which $w_1w_2$ has
  constant phase are the $\alpha$-planes.}  Taking a parallel
propagated frame of $E'$ from $i^-$ up the generators of $\scri^-$
will therefore yield a gauge in which $ A ^-= A ^-(w_1, \eta)(\d \eta
+iw_1\d\bar w_1 -i\bar w_1 \d w_1)$ and this $ A ^-(w_1,\eta)$ will be
a smooth function on $\scri^-$ with values in $\su(n)$ that vanishes
at $i^-$.  We must also require that the holonomy of $A^-$ about the
unit circle in each $w_1$ plane (of fixed $\eta$) should vanish as
these circles bound $\alpha$--planes in $\M^+$ on which the connection
must be flat. The function $A^-(w_1,\eta)$ is otherwise freely
prescribable.  To avoid some technicalities, we will, however, assume
that $ A ^-(w_1,\eta)$ vanishes at $\Sigma\cap\scri$ (i.e., at
$|w_1|=1$) and near $i^-$.\footnote{Both conditions are stronger than
  we need.  The first guarantees the vanishing of the holonomy around
  $|w_1|=1$ and makes certain choices we will make later canonical.
  The second condition guarantees that $A^-$ is the restriction of a
  smooth 1-form from $\widetilde\M$ at $i^-$.}
%  The restriction of a smooth function to
%the vertex of a cone has restrictions on the degree of spherical
%harmonics that are admissable in each order of the expansion in a
%Taylor series away from the vertex $i^-$, Penrose (1963).  The
%following discussion will apply to data $ A ^-$ that only vanishes
%only at $i^-$ as opposed to on an open set near $i^-$; in general the
%corresponding solutions would have some degree of irregularity at
%$i^\pm$, but would otherwise be smooth on $\M^+$ and $\scri^\pm$.}
Our aim is to show that $A^-$  gives rise to a unique solution to the
ASDYM equations on $\M^+$ and find a procedure to determine the
`final' data on $\scri^+$.  

%The antipodal map on $\widetilde \M$ sends
%$\scri$ to itself giving a canonical identification between $\scri^+$
%and $\scri^-$ (the light-cone of a point of $\scri^-$ reconverges on
%the `antipodal' point on $\scri^+$) and so we can directly compare
%initial data on $\scri^-$ to final data on $\scri^+$ to see whether
%the scattering is non-trivial.

\subsection{$\beta$-planes and their twistor theory} 
%We will derive the scattering on each $\beta$-plane on $\scri$
%separately; there is no interaction between the data on distinct
%$\beta$-planes.  
The $\beta$-planes on $\scri$ are given by
$w_1=\e^{i\eta}w_2$ and denoted $W'_\eta$; they foliate $\scri-i^\pm$.
All $W'_\eta$ intersect at $i^\pm$ but are otherwise disjoint and are
topologically 2-spheres with complex stereographic coordinate $w_1$.
The hypersurface $\Sigma$ cuts $W'_\eta$ into the two discs
$W'^\pm_\eta=\{W'_\eta, \pm(|w_1|- 1)\geq 0\}$.  
%We will restrict the
%given initial data to each $W_\eta'^-$ and generate the associated
%final data on each $W_\eta'^+$.

The correspondence with twistor space $\PT$ is as follows.  The
points $i^\pm$ correspond in twistor space to a real line $I$
(the $\pm$ corresponding to the choices of orientation of $I$).  Each
$\beta$-plane $W'_\eta$ on $\scri$ is dual to a 2-plane
$W_\eta\subset\PT$ that contains $I$.  Given $Z\in\PT-I$, the
corresponding $\alpha$-plane $\alpha_Z\subset \widetilde\M$ intersects
$\scri$ in a null geodesic that lies in a unique $W'_\eta$ as a great
circle $Z'=\alpha_Z\cap W'_\eta(=\alpha_Z\cap\scri)$ and we define $W_\eta$
to consist of those $Z$ such that $\alpha_Z\cap\scri\subset W'_\eta$.  A point
$p\in W'_\eta$ corresponds to the line in the projective plane
$W_\eta$ consisting of those $Z'$ through $p$; thus $W_\eta$ and
$W'_\eta$ modulo the antipodal map (which reduces $W'_\eta$ from $S^2$
to $\RP^2$) are in projective duality. 
% and this line gives a great circle in the 2-sphere obtained as the
%double cover $\widetilde W_\eta$ of $W_\eta$.

We can realise this in coordinates as follows.  Given homogenous
coordinates $z_i$, $i=1,2,3$ on $W_\eta$, the corresponding null
geodesic on $W'_\eta$ can be taken to be the great circle represented
in terms of the stereographic coordinate $w_1$ as $$z_3 (1-|w_1|^2) +
\Re ((z_1+iz_2) w_1)=0.$$
This gives $I$ as the line $z_3=0$.  The
intersection $\Sigma_\eta=\Sigma\cap W'_\eta=\p W'^\pm_\eta$ is a null
geodesic corresponding to the point $Z_{\Sigma_\eta}\in W_\eta$ with
homogeneous coordinates $(0,0,1)$.  Thus, a general real twistor
$Z\in\PT_\R-I$ can be parametrized by $Z=(z_i,\eta)$ where $z_i\sim
\lambda z_i, \lambda\in\R^*$ (globally this is a blowup of $I$ in
$\PT_\R$). 

Let $\widetilde W_\eta$ be the double cover of $W_\eta$.  It can be
represented as the unit sphere $\sum z_i^2=1$ and is in duality with
the sphere $W'_\eta$ in the sense that points of one corresponds to
oriented great circles in the other.

\subsection{The scattering on $W'_\eta$}
In the following, we fix a value of $\eta$, and all our considerations
will be concerned with the relationship between initial data $A^-$ on
$W'^-_\eta$, a $\U(n)$ valued matrix function $h(z_i,\eta)$ on
$\widetilde W_\eta$, twistor data $H(z_i,\eta)$ on $W_\eta$ and then
final data on $W'^+_\eta$.

Given the initial data consisting of the connection $A^-$ restricted
to $W'^-_\eta$, we first fix a covariantly constant frame $f:E'=\C^n$
of the Yang-Mills vector bundle $E'$ around $\p W'^-_\eta$ (which
exists as $A^-$ has  trivial holonomy  around $\p
W'^-_\eta$).  
\begin{defn}\label{holdef}
We define $h(z_i,\eta):\widetilde W_\eta\rightarrow
\U(n)$ to be the holonomy around the loop formed by the oriented
geodesic in $W'^-_\eta$ corresponding to $(z_i,\eta)\in\widetilde
W_\eta$ and one of the arcs of $\p W'^-_\eta$ required to close the
semicircle (because of the vanishing of holonomy around $\p
W'^-_\eta$, it doesnt matter which arc is chosen).
\end{defn}
For $\widetilde Z\in \widetilde W_\eta$, a change in orientation of
the corresponding geodesic in $W'_\eta$ corresponds to $\widetilde
Z\rightarrow -\widetilde Z$ and  we will therefore have $h(-\widetilde
Z')=h(\widetilde Z')^{-1}$.  We also note that as $\widetilde Z'$
tends towards a semicircle of $\p W'^-_\eta$, i.e.,
$z_1,z_2\rightarrow 0$, $h\rightarrow I_n$.
\begin{defn} We now define the twistor data $H:\PT_\R\rightarrow
  \Herm_n$, $\Herm_n$ being the $n\times n$ hermitian matrices, to be
  \be{Hdef} H:=g^{-*}g^-\, \ee where $g_-$ is the solution to the
  following Birkhoff factorization problem in the $z_3$ variable:
  \be{hfact} h(z_i, \eta)g^-(z_i, \eta)= g^+(z_i,\eta)\, . \ee
\end{defn}
Here we have fixed a positive scaling of the homogenous coordinates by
setting $|z_1^2+z_2^2|=1$ and $g^\pm$ extends holomorphically over
$\pm\Im z_3\geq 0$ in the complex $z_3$ plane for each real $z_1,
z_2$, and we normalize $g^\pm$ by the condition that, at $z_3=\infty$,
$h=g^\pm=I_n$.  $H$ is clearly a positive definite Hermitian matrix
function of $(z_i,\eta)$.

\smallskip

\noindent
{\bf Remarks:} 1. This is where we use the smallness assumption on the
data: the Birkhoff factorization in general exists only when we also
allow a factor of $\Delta$ with diagonal entries given by powers of
$(z_3+i)/(z_3-i)$, but $\Delta$ is the identity matrix for
sufficiently small $h$.

\noindent
2. We have also used the vanishing of $A^-$ on $\Sigma\cap\scri$ in the
   normalization conditions for $g^\pm$.  Had we not done so, the
   normalization condition would be more complicated.

\begin{lemma}
$H(-\widetilde Z)=H(\widetilde Z)$ so that $H$ is defined
on $\PT_\R$.
\end{lemma}
\proof We first note that $g^{-*}g^-=g^{+*}g^+$ follows from the
unitarity of $h$.  Furthermore, $h(-\widetilde
Z)=h(\widetilde Z)^{-1}$ implies that $h(-\widetilde
Z)^{-1}g^-(\widetilde Z)=g^+(\widetilde Z)$ so that $g(-\widetilde
Z)^\pm= g(\widetilde Z)^\mp$ by the uniqueness of the Birkhoff
factorization.  Thus \be{Hprops} H(-\widetilde
Z)=g^{+*}g^+=g^{-*}g^-=H(\widetilde Z)\ee as required. $\Box$

\smallskip

This $H$ can now be used as in the main theorems to determine an ASDYM
field on $\widetilde\M$.  We need to show that it correctly reproduces
the initial data on $\scri^-$.  Clearly we only need to see that it
correctly reproduces the appropriate initial data on each $W'^-_\eta$.
It is sufficient to prove that the connection that it determines on
$W'^-_\eta$ leads to the given holonomy matrix $h(z_i,\eta)$ because
of the following theorem:

\begin{thm}[Manakov \&
Zakharov 1981, Novikov 2002]\label{novthm} Let $A$ be a twice differentiable $\U(n)$
connection on the projective plane $\RP^2$ that vanishes on the line
at $\infty$, $l_\infty$ and let $h$ be the $\U(n)$-valued function on
the space of oriented lines $S^2$ in $\RP^{2}$ obtained by expressing
the holonomy around each line in a fixed covariantly constant
trivialisation of the bundle over $l_\infty$.  Then if $A$ is
sufficiently small, then $h$ determines $A$ uniquely up to gauge
transformations.
\end{thm}

The key ingredients of the proof of this theorem are
contained in the Birkhoff factorization of equation 
(\ref{hfact}), the definition of $H$ in equation (\ref{Hdef}) and the
reconstruction of the connection from $H$ when $E$ is trivial of
Theorem \ref{mainthm}.

We note that the theorem in Novikov (2002) is actually expressed in
terms of connections on $\R^2$ with suitable fall-of conditions, but
they are easily seen to be slightly weaker than the statement above.
This applies to our situation by considering $W_\eta$ to be a double
cover of a projective plane $\RP^2$ using the joint antipodal map on
$S^2\times S^2$.  Under this map, great circles map to lines and $\p
W'^-_\eta$ can be taken to be a double cover of the `line at
$\infty$', $l_\infty$; $W'^-_\eta-\p W'^-_\eta$ then maps $1:1$ to
$\RP^2-l_\infty$.

It remains to prove that the connection that $H$ gives rise to on
$\scri^-$ has holonomy given
by $h$.  This in fact is a special case of a more general proposition which
is informative in its own right: 

\begin{propn}
Consider an ASDYM field on $\widetilde\M$ for which the twistor data
has trivial vector bundle $E$ and is determined by a Hermitian metric
$H$ on $E|_{\PT_\R}$.  A pair $Z_0, Z$ of $\alpha$-planes in
$\widetilde\M$ intersect in two antipodal points $x^\pm$ corresponding
to the two orientations of the line $L$ in $\PT_\R$ joining $Z_0$ and
$Z$.  Let $l$ be a loop in $Z_0\cup Z$ going from $x^-$ to
$x^+$ in $Z_0$ and returning in $Z$.  In a fixed covariantly constant
frame on $Z_0$, $h_l$ is related to $H|_L$ by \be{holfact}
h_l=g(x^+,Z)g(x^-,Z)^{-1}\, , \ee where $g(x^\pm,Z)$ are the solutions
to the Birkhoff factorizations \be{mainfactagain}
H=g^*(x^\pm,Z)g(x^\pm,Z) \ee such that $g(x^\pm,Z)$ are respectively
holmorphic on the two distinct holomorphic discs $D_{x^\pm}\subset
\PT$ with boundary $L$ whose union is the complex projective line $\C
L$ obtained by complexifying $L$.
\end{propn}
Note that the holonomy $h_l$ of the ASDYM connection depends only on
the topology of $l$ because the connection is flat on $\alpha$-planes.
\smallskip

\noindent {\bf Proof:}
%The two points $x^\pm\in\widetilde\M$ correspond to two holomorphic
%discs $D_{x^\pm}$ in $\PT$ with common boundary $L^\pm$ and whose
%union is a projective line in $\PT$.  
In the construction of $(E,H)$ on $\PT$ from an ASDYM field on
$\widetilde\M$, a frame of $E$ at a point $Z\in\PT_\R$ has the
interpretation as a covariantly constant frame for the Yang-Mills
bundle over the corresponding $\alpha$-plane $Z'$ in $\widetilde\M$.
On the other hand, in the factorization problem $H=g^*(x,Z)g(x,Z)$
where $g$ is defined holomorphically for $Z\in D_x$, $g$ has the
interpretation as a map from a frame of $E_Z$ (i.e., covariantly
constant on $\alpha_Z$ to a unitary frame of $E'_x$.  Putting these
facts together, we see that, if we normalize
$H(Z_0)=g(x^\pm,Z_0)=I_r$, in a given unitary frame for $E_{Z_0}$,
then $h_l$ as given in equation (\ref{holfact}) is the product of the
map from a covariantly constant frame along $Z_0$ to a covariantly
constant frame along $Z$ at $x^-$ with the map from the covariantly
constant frame along $Z$ at $x^+$ to that on $Z_0$ at $x^+$ and is
hence the holonomy as required.  $\Box$

\begin{corol}If $H$ is defined according to equations (\ref{Hdef}) and
  (\ref{hfact}), then the connection that $H$ gives rise to on
$\scri^-$ has holonomy, as in definition \ref{holdef} equal to
the holonomy from which it was obtained.
\end{corol}
{\bf Proof:} We wish to calculate the holonomies $h(\widetilde Z)$ of
the Yang-Mills connection around the loops in $W'^-_\eta$ obtained by
joining $\widetilde Z'$ to one of the components of $\p W'^-_\eta$.
These should be evaluated in a frame that is covariantly constant
around $\p W'^-_\eta$.  This is a special case of the above
proposition as can be seen by taking
$Z_0=Z_{\sigma_\eta}$ and $Z\in W_\eta$ so that $g(x^\pm,Z)$ can
be identified with $g^\pm$ of equations (\ref{hfact},\ref{Hdef}) and
(\ref{Hprops}).  Thus, given $H$, equations (\ref{Hprops}) and
(\ref{Hdef}) can be considered to be the Birkhoff factorizations that
give rise to a solution whose holonomy $h(z_i, \eta)$ is determined by
(\ref{hfact}) and hence gives the original $h(z_i,\eta)$ as required.
$\Box$

\subsection{Summary}
The scattering map from initial data $ A _-$ on $\scri^-$ to final
data $ A ^+$ on $\scri^+$ can be constructed by first constructing
a family of parallel propagators $h(\widetilde Z):\widetilde\PT_\R\rightarrow
\U(n)$ along the intersections of the $\alpha$-planes with $\scri^-$;
$h$ is defined on
the double cover of twistor space.  This can be used to find a
hermitian metric $H$ on the restriction of a trivial bundle $E$ to
$\PT_\R$ via the Birkhoff factorization problems (\ref{hfact},
\ref{Hdef}).  Finally, $H$ can be used to recover an ASDYM field on
$\widetilde \M$ and in particular final data $ A ^+$ on $\scri^+$
via the proof of theorem \ref{mainthm}.
each of these maps is $1:1$ and onto assuming small data.

\subsection{The nontriviality of the scattering}
We note that if we take theorem (\ref{novthm}) as giving an equivalence
between connections $ A ^-$ on $\scri^-$ and holonomy data
$h(\widetilde Z)$ on $\widetilde\PT_\R$, then the final data $ A ^+$
is simply encoded in the holonomy data $h^{-1}(\widetilde Z)$.  This
follows directly because the total holonomy around a curve in an
$\alpha$-plane must be the identity as the curvature of an ASDYM
connection is zero on $\alpha$-planes.  In these terms, the scattering
might seem rather trivial.  In particular for $\U(1)$ connections we
can see directly that there exists a gauge in which the
scattering map is just reversal of sign: $ A ^+ = -\tilde\sigma^* A
^-$ where $\tilde\sigma$ is the antipodal map.  However, for
non-abelian connections, there is no such simple formula relating $ A
^+$ to $ A ^-$ and the scattering at the level of the connection will
be quite non-trivial.  An appropriate geometrical analogue to consider
here is the case of a Zoll surface which is a topological 2-sphere
with a metric whose geodesics are all closed.  Although Zoll
perturbations of the round metric correspond to variations of the
conformal factor that are odd under the antipodal map, finite Zoll
perturbations are highly nontrivial and can only be completely
characterised in the axisymmetric case, see for example LeBrun \&
Mason (2002).

It is worth noting furthermore that theorem (\ref{novthm}) is proved
using the combination of Birkhoff factorizations above and these are
non-trivially different for antipodal points $x^+\in \scri^+$ and
$x^-\in\scri^-$.  The twistor data giving rise to the fields at
$x^\pm$ is simply $H$ restricted to the corresponding line $L_{x^\pm}$
which is the same line with the $\pm$ determining opposite
orientations.  The discs $D_{x^\pm}$ meet at $L_{x^\pm}$ and make up
a complex line.  To construct the ASDYM field at $x^\pm$, we solve the
Birkhoff factorization problems \be{antipodfact} H= g^\pm g^{*\pm}\ee
where $g^\pm$ extend holomorphically over $D_{x^\pm}$ respectively.
Since $D_{x^\pm}$ are complex conjugate discs in $\PT$, $g^{*\pm}$
extend holomorphically over $D_{x^\mp}$ respectively.  Thus the two
factorization problems in equation (\ref{antipodfact}) are in opposite
orders and will have non-trivially distinct solutions in the
non-abelian case.

\section{Further developments}\label{conclusions}
There are a number of questions that remain.  

One problem is to find interesting examples of the different types of
ASDYM fields in split signature.  We have constructed here solutions
with $c_1\neq 0$ but $c_2=c_3=0$, and in the appendix we have an
example with $c_2=2$, $c_1=c_3=0$, but it would be interesting to have
also an example with non-trivial $c_3$ or $\alpha$-invariant, but
$c_1=c_2=0$.  We also have not produced non-abelian examples with
trivial $E$, but non-trivial $H$---such a solution would explicitly
demonstrate the non-triviality of the scattering.

Perhaps the most interesting questions relate to the relationship
between the constructions above and twistor-string theory.  In
particular, the version of twistor-string theory due to Berkovits, see
for example Berkovits \& Witten (2004), focuses on curves in $\PT$
with boundary on $\PT_\R$.  It would be intriguing to know how the
constructions of this paper sit as an ASD sector inside those of
Berkowitz which apply to the full Yang-Mills equations.

%Scattering of ASD fields in Minkowski signsture is well
%known to be trivial perturbatively, but this is not the case in the
%complex, nor in split signature.  It would be interesting to
%understand the relationship between the algorithm for calculating
%scattering given above with that arising from twistor-string theory in
%Witten (2004).

%There has already been work
%relating the Ward transform to twistor-string theory, see Lechtenfeld
%and Popov (2004) and Popov \& S\"amann (2004).

\appendix
\section{Proofs for theorem \ref{mainthm}}\label{proofs}

We first prove
\medskip

\noindent
{\bf Lemma (\ref{techlemma}):} The given definition of $E\rightarrow
\PT_\R$ is a smooth extension of $E$ from $\PT-\PT_\R$ such that the
d-bar operator $\bar\p_E$ extends smoothly over $\PT_\R$.

\medskip

\proof
Choose a small neighbourhood $U$ of $Z\in\PT_\R$ and a smooth frame 
$f$ of $E$ over $U_\R=U\cap\PT_\R$.  This determines a smooth frame 
$q^*f$ of $E'$ over $q^{-1}(U_\R)$ that is covariantly constant along
the foliation of $q^{-1}(U_\R)$ by horizontal lifts of
$\alpha$-planes.  We extend this frame smoothly over $q^{-1} U$ as
follows:
\begin{itemize}
\item We first construct a formal power series on $q^{-1}(U_\R)$ in a
  direction transverse to $q^{-1}(U_\R)$ in $q^{-1}U$ by requiring that
  the formal power series be holomorphic to all orders on the fibres
  of $p$ in $q^{-1}U\subset \F$.  Each fibre of $p$ can be expressed
  as an upper-half plane $H$ on which $f$ is defined on some interval
  on the real axis.  If $z=x+iy$ is a holomorphic coordinate on $H$
  (say $z=\pi_{1'}/\pi_{0'}$ in some affine coordinate system as above),
  then the condition $\p^k f/\p\bar z^k=0$ for all $k$ determines
  $\p^kf/\p y^k$ uniquely for all $k$ inductively in terms of $\p^l
  f/\p x^l$ for $l\leq k$.  Thus the formal power series is defined
  uniquely and hence globally.
  
\item Using Borel's lemma, we can take a smooth extension of $f$ to
  $q^{-1}U$ whose power series on $q^{-1}(U_\R)$ is the given formal
  power series, and this can be made global by use of a partition of
  unity. 
\end{itemize}
In the frame thus constructed, we have that the d-bar operator on
twistor space vanishes to all orders at $\PT_\R$, and hence extends
smoothly over $\PT_\R$.  This can be seen as follows.

The d-bar operator is given by
\begin{eqnarray*}
\bar\p_E&=&\d\bar\pi_{A'}\frac{\p}{\p\bar\pi_{A'}} +
\frac1{\pi^{A'}\bar\pi_{A'}}\d x^{AA'}\bar\pi_{A'}\pi^{B'}\p_{AB'}
\\ && + f^{-1}\left( \d\bar\pi_{A'}\frac{\p}{\p\bar\pi_{A'}} +
\frac1{\pi^{A'}\bar\pi_{A'}}\d x^{AA'}\bar\pi_{A'}\pi^{B'}\D_{AB'}\right)f
\end{eqnarray*}
where $D_{AA'}$ is the given Yang-Mills connection.  We claim that
the potentially singular latter term $f^{-1}\left(
  \d\bar\pi_{A'}\frac{\p}{\p\bar\pi_{A'}} +
  \frac1{\pi^{A'}\bar\pi_{A'}}\d
  x^{AA'}\bar\pi_{A'}\pi^{B'}\nabla_{AB'}\right)f$ vanishes at
$\pi^{A'}\bar\pi_{A'}=0$ to all orders by construction.  Firstly, the
  fact that 
$\p
f/\p\bar\pi_{A'}$ 
vanishes to all orders at $\pi^{A'}\bar\pi_{A'}=0$
follows immediately from the construction. 

By definition of $E|_{\PT_\R}$ ,  $\pi^{A'}D_{AA'}f=0$ when $\pi_{A'}$
is real.  Introducing an affine coordinate
$z=(\pi_{0'}+i\pi_{1'})/(\pi_{0'}-i\pi_{1'})$ on the fibres of $\F$ (so
that $\F_\R$ is given by $|z|=1$) the vanishing of $\p
f/\p\bar\pi_{A'}$ to all orders is equivalent to $\p^kf/\p\bar z^k$
for all $k$.  Furthermore
$$
(\p/\p\bar z)^k (\pi^{A'}D_{AA'}f)=
\pi^{A'}D_{AA'}((\p/\p\bar z)^k f)=
0
$$ so that, by uniqueness of the extension, $\pi^{A'}D_{AA'}f=0$ to
all orders on $q^{-1}U_\R$.  Hence the latter, potentially singular
terms, vanish to all orders at $U_\R$ and hence extend smoothly over
$U_\R$ as desired.$\Box$

\section{Kahler structure and explicit solutions}
We develop a formalism for the correspondence in the case that a
complex structure on $\widetilde \M$ is chosen.  This reduces the
symmetry group of twistor space from $\PSL(4,\R)$ to $\PSO(4)$.  The
affine coordinates of \S\ref{standardcoords} respect the choice of a
`Lorentz' $\SO(2,2)$ subgroup of the space-time conformal group
$\PSO(3,3)$, but here we will focus instead on a formalism that is
invariant under the $\SO(3)\times\SO(3)$ subgroup of rigid rotations
of each of the $S^2$ factors.  This formalism is particularly well
adapted to giving explicit descriptions of certain exact solutions
which are described after the next subsection.

\subsection{The pseudo-kahler correspondence}
The space $\widetilde\M$ admits complex structures for which the
metric (\ref{s2xs2}) is scalar-flat pseudo-kahler.  The complex
structure realizes $\widetilde\M$ as the complex manifold
$\CP^1\times\CP^1$ and via the representation of $\PT-\PT_\R$ as the
bundle of metric and orientation compatible complex structures over
$\widetilde\M$, naturally embeds 
$\widetilde\M$ into $\PT$ as a quadric that misses $\PT_\R$.

On choosing a positive definite quadratic form $Q$ on $\T_\R$, we
develop an $\SO(4)$ invariant formalism.  Let $Z^{\alpha \dot\alpha}$
be coordinates on $\T$ where $\alpha =0,1$ and $\dot\alpha=\dot 0,\dot
1$ are $\SU(2)$ spinor indices for $\SO(4)=\SU(2)\times\SU(2)/\Z_2$
(we use different indices to avoid confusion with the $\SL(2,\R)$
spinor indices used above for $\SO(2,2)$).  Recall that for $\SU(2)$
spinors we raise and lower indices with the alternating spinors
$\varepsilon_{\alpha \beta }=-\varepsilon_{\beta \alpha}$,
$\varepsilon_{\dot\alpha\dot\beta}= -
\varepsilon_{\dot\beta\dot\alpha}$,
$\varepsilon_{01}=\varepsilon_{\dot 0\dot 1}=1$.  We also have the
quaternionic complex conjugation $x^\alpha\rightarrow \hat
x^\alpha=(-\bar x^1, \bar x^0)$ so that $\T_\R$ consists of those
$Z^{\alpha\dot\alpha}$ such that $\hat
Z^{\alpha\dot\alpha}=Z^{\alpha\dot\alpha}$.

Points of $\widetilde \M$ can be identified with points of the quadric
$Q\subset\PT$ (using an obvious abuse of notation)
$$
Q=\{Z\in\PT| Q:=Z^{\alpha\dot\alpha}Z_{\alpha\dot\alpha}=0\}.$$  This
gives $Z^{\alpha\dot\alpha}=x^\alpha  y^{\dot\alpha}$
on $Q$ and each of $x^\alpha$ and $y^{\dot\alpha}$ can be thought of
as homogeneous 
coordinates on the $\CP^1$ factors of $\widetilde\M=\CP^1\times \CP^1$
as a complex manifold.

We take $(x^\alpha,y^{\dot\alpha})$ to be homogeneous coordinates on
$\widetilde\M$ normalized so that $x^\alpha \hat x_\alpha=y^\alpha\hat
y_\alpha= 1$ related to the previous affine coordinates by
$w_1=x^1/x^0$ and $w_2=y^{\dot1}/y^{\dot0}$.  Homogeneous functions
$f(x^\alpha,y^\alpha)$ will be said to have weight $(p,q)$ if
$f(\e^{i\theta}x^\alpha, \e^{i\phi}y^\alpha)=\e^{(p i\theta+q
i\phi)}f(x^\alpha, y^\alpha)$.  Such homogeneous functions can also be
taken to represent sections of the tensor product of $p$ copies of the
spin bundle on the first $S^2$ factor with $q$ copies of the spin
bundle on the second factor.

We can introduce homogenous coordinates
$(x^\alpha,y^\alpha,\lambda_A)$ on $\F$ where
$\lambda=\lambda_1/\lambda_0$, $|\lambda|\leq 1$ is a coordinate on
the unit disc $D_x$, and $\lambda_0$ has weight $(-1,-1)$, $\lambda_1$
has weight $(1,1)$ so that $\lambda$ has weight $(2,2)$ and the
incidence relation is
$$
Z^{\alpha\dot\alpha}=\lambda_0 x^\alpha  y^{\dot\alpha}+\lambda_1
\hat x^\alpha \hat y^{\dot\alpha} \, .
$$
This can be expressed implicitly as
$$
Z_{\alpha\dot\alpha}x^\alpha \hat y^{\dot\alpha}=Z_{\alpha\dot\alpha}\hat
x^\alpha y^{\dot\alpha}=0
$$
which can be seen to reduce to one complex equation when
$Z^{\alpha\dot\alpha}$ is real yielding the formula $x^\alpha\propto
Z^\alpha_{\dot\alpha}\hat y^{\dot\alpha}$ for real $\alpha$-planes,
i.e., a real $\alpha$-plane is the graph of an orientation reversing
isometry from the second to the first factor.

In these coordinates the Lax pair can be expressed in terms of the
holomorphic and antiholomorphic exterior derivatives $\eth_x, \bar
\eth_x$ and $\eth_y, \bar\eth_y$ on the $S^2$ factors with homogeneous
coordinates $x^\alpha$, $y^{\dot\alpha}$ respectively.
 Rather than taking these as form valued operators we
can instead can take them to have weights $(-2,0), (2,0), (0,-2)$ and
$(0,2)$ respectively.  We have
$$
L_0= \lambda\eth_x-\bar\eth_y\, , \qquad L_1= \lambda\eth_y-\bar\eth_x\,
. 
$$  
$L_0$ and $L_1$ define the complex structure on the complement of $\p
\F$ and are tangent to the fibration $\p \F\rightarrow \PT_\R$.

\subsection{ADHM description of instantons with $c_2=2$}
Since the split-signature instantons correspond to stable bundles on
$\PT$ they can be constructed using an adaptation of the ADHM
construction to split signature.  The ADHM construction expresses a
rank $n$
bundle $E\rightarrow \PT$ with $c_2=k$  as the cohomology of the sequence
$$
V\stackrel{K\cdot Z}\longrightarrow W\stackrel{K^*\cdot
  Z}\longrightarrow \bar V^*\, .
$$
Here $V,W$ are complex vector space of dimension $k$ and $2k+n$
respectively, $W$ has a pseudo-Hermitian metric $h$ (which is used to
deine the $*$ operation below) and $K:V\otimes \T\rightarrow W$ are
linear maps such that $K^*\cdot Z\circ K\cdot Z=0$ for all $Z$.  All
bundles corresponding to instantons arise in this way.  Given the ADHM
data of the matrices $K$, the corresponding solution on
space-time can be written down explicitly, Atiyah (1979) and will be
non-singular if certain non-degeneracy conditions are satisfied.

Here we work through the first non-trivial case of $k=n=2$.  In this
case we can choose our frames of $\T$, $V$ and $W$ so that coordinates
on these spaces have the index structure $Z^{\alpha \dot\alpha}$ as
above, $v^\alpha $ and $w^{\alpha _1\alpha _2\dot\alpha}= w^{(\alpha
  _1\alpha _2)\dot\alpha}$ respectively. In this frame the map $K$ is
$K_{\alpha \beta \dot\beta}^{\gamma _1\gamma _2\dot\gamma
}=\delta_\alpha ^{(\gamma _1}\delta_\beta ^{\gamma
  _2)}\delta_{\dot\beta}^{\dot\gamma }$ and the hermitian metric on
$W$ is
$$
h(w,w)=w^{\alpha _1\alpha _2\dot\alpha}\hat w^{\beta _1\beta
  _2\dot\beta}h_{\alpha _1\alpha _2\beta _1\beta
  _2}i\varepsilon_{\dot\alpha\dot\beta}\, ,
$$
where $h_{\alpha \beta \gamma \delta }=h_{(\alpha \beta \gamma
  \delta )}=\hat h_{\alpha \beta \gamma \delta }$ and $v^\alpha
\rightarrow \hat v^\alpha $ is the standard quaternionic $\SU(2)$
conjugation. With this, the ADHM equations $K^*\cdot Z\circ K\cdot
Z=0$ are satisfied.

In order to write down the corresponding solution on space-time, it is
convenient to use the representation of space-time above as the
quadric $Q$.  A point $x\in\M$ is represented by the pair of spinors
$(x^\alpha , y^{\dot\alpha})$ which correspond to the line joining
$Z^{\alpha \dot\alpha}_1=x^\alpha y^{\dot\alpha}$ and $Z_2^{\alpha
  \dot\alpha}=\hat x^\alpha \hat y^{\dot\alpha}$ in $\PT$.  The fibre
$E'_x$ of $E'\rightarrow \M$ at $x$ can be represented as the subspace
of $W$ in the kernel of $K^*\cdot Z_1$ and $K^*\cdot Z_2$.  The
projector $P_x:W\rightarrow W$ onto the subspace $E'_x$ can be
constructed as
$$
P_x=I-K\cdot Z_1\Delta(x)^{-1}K^*\cdot Z_2 + K\cdot Z_2\Delta(x)^{-1}
K^*\cdot Z_1 \,
$$ 
where 
$$
\Delta(x):=K^*\cdot Z_2\circ
K\cdot Z_1= -K^*\cdot Z_1\circ
K\cdot Z_2
$$ and the latter identity follows from the ADHM equation.  A smooth
(but not holomorphic) unitary frame for $E'_x\subset W$ is given
explicitly by
$$
%w^{\alpha_1\alpha_2\dot \alpha}= a 
U=\frac1{\sqrt{h^{-1\alpha_1\alpha_2\beta_1\beta_2
    }x_{\alpha_1}x_{\alpha_2}\hat x_{\beta_1}\hat x_{\beta_2}}} \{
y^{\dot\alpha} h^{-1\alpha_1\alpha_2\beta_1\beta_2 }\hat x_{\beta_1}\hat x_{\beta_2},
\hat  y^{\dot\alpha}
h^{-1\alpha_1\alpha_2\beta_1\beta_2 }x_{\beta_1} x_{\beta_2}\}
$$

The connection is given by projecting infinitesimally $E_x$ to
$E_{x+\delta x}$ inside $W$, thus
$$
\nabla U=P_x \d U \, ,
$$
which can now be calculated explicitly by the reader with the energy
and inclination.

We note that although a formalism based on the double cover
$\widetilde\M$ of $\M$ has been used here, the formula for $P_x$ is
invariant under the antipodal map and the solution descends to $\M$.

The construction breaks down when $\Delta(x)=x^\gamma \hat x^\delta
h_{\gamma \delta \alpha \beta }$ is singular.  The determinant of
$\Delta(x)$ is $d=x^\gamma \hat x^\delta h_{\gamma \delta \alpha \beta
}x^\epsilon\hat x^\phi h_{\epsilon\phi}^{\alpha \beta }$ and so
$\Delta$ is non-degenerate for all $x$ if $h$ is non-degenerate when
regarded as a symmetric trace-free $3\times 3$ matrix over spinors
$V^{\alpha \beta }=v^{(\alpha \beta )}$.  (It is worth noting that the
locus $d=0$ in the complex is the same as that defining the jumping
lines as described in \S\ref{2inst} given by $S(\bx,\bx)=\bx\cdot\bx$
where $S-1$ is identified with a constant multiple of
$h_{\alpha_1\alpha_2
\gamma_1\gamma_2}h^{\gamma_1\gamma_2}_{\beta_1\beta_2}$ where the
symmetric pairs of spinor indices $\alpha_1\alpha_2$ are identified
with 3-vector indices in the standard way; it can be checked that the
condition that $S$ arises from a traceless spinor in this way is
equivalent to the Poncelet condition $\tr S^2=3/2$.)

\subsection{Yang's J-matrix formulation}\label{J}
On a (pseudo-)kahler 4-manifold, the ASDYM equations can be recast as
the condition that the bundle with unitary connection $(E',D)$ be
compatible with the complex structure and satisfy in addition the
condition that $\omega\wedge F=0$ where $\omega$ is the Kahler 2-form
and $F$ is the curvature of the connection.  Given a holomorphic
vector bundle, Chern's theorem states that unitary connections
compatible with the given complex structure are in $1:1$
correspondence with hermitian metrics $J$ on $E'$.  In a local
holomorphic frame, the connection on the bundle is obtained by
differentiation of $J$ and the ASDYM equation is given by
$\omega\wedge \bar\p (J^{-1}\p J)=0$.  In the physics literature, $J$
has become known as Yang's J-matrix.

In the simplest situation, $E'$ will be trivial as a holomorphic
vector bundle over $\widetilde\M=\CP^1\times\CP^1$ and so the
holomorphic frame will be defined globally up to constant $\GL(n,\C)$
transformations.  Thus so will $J$.  

If $E$ and $E'$ are both trivial, then $J$ and $H$ will be related by 
$J(x^\alpha,y^{\dot\alpha})=g(x^\alpha y^{\dot\alpha})^{-1}g(x^\alpha
y^{\dot\alpha})^{*-1} 
$
where $g$ is as defined in equation (\ref{mainfact}).

\subsection{Explicit solutions from Ward ansatze}\label{kwsoln}
We consider the ansatze of \S\ref{ward} and work throught the
procedure to obtain the $J$-matrix of the bundle on $\widetilde\M$.
We note that, as a holomorphic vector bundle over $Q$, $E'$ is
non-trivial since it is the restriction of
$E=\Oc(1)\oplus\Oc(-1)$ to $Q$ from $\PT$.  On $Q$,
$E'=E|_Q=\Oc(1,1)\oplus\Oc(-1,-1)$ where $\Oc(p,q)$ is defined to be the
tensor product of the pullback of $\Oc(p)$ from the first $\CP^1$
factor (coordinatised by $x^\alpha $) with $\Oc(q)$ from the second one
(coordinatised by $y^{\dot\alpha}$).  Thus, the $J$ matrix can only be
presented globally if its entries are understood to take values in the
appropriate line bundles.

In order to make clear the holomorphic nature of these line bundles we
do not in the following normalize $x^\alpha \hat x_\alpha =1$ etc..
We also use homogeneous coordinates $(\lambda_0,\lambda_1)$ on the
discs $D_x$ so that $\lambda=\lambda_1/\lambda_0$ and the incidence
relation becomes
$$
Z^{\alpha \dot\alpha}=\lambda_0 x^\alpha y^{\dot\alpha}+\lambda_1\hat
x^\alpha \hat y^{\dot\alpha}\, . 
$$
Note that when $Z^{\alpha \dot\alpha}$ is real, $\lambda\bar\lambda=1$,
i.e., $\bar\lambda=1/\lambda$.  In terms of homogeneous coordinates
we have
$(\lambda_0,\lambda_1)=(\bar\lambda_1,\bar\lambda_0)$.

We consider the adaptation of the Ward ansatze given in \S\ref{ward}
in the case that $k=1$ and $Q$ is the quadric as above.  We will take
$f$ to be an arbitrary smooth function on $\PT_\R$ corresponding to an
arbitrary smooth solution $\phi$ to the ultra-hyperbolic wave equation
on $\M$ by means of the X-ray transform of the function $f/Q$ which in
these coordinates becomes \be{phi} \phi(x)=\frac1{2\pi
i}\oint_{|\lambda|=1} f(x^\alpha y^{\dot\alpha}+\lambda \hat x^\alpha
\hat y^{\dot\alpha})\frac{\d \lambda}{ \lambda} \, .  \ee

An intermediate step in finding the solutions in the abelian case with
patching function is $\e^f$ is to find a function $g(x,\lambda)$
holomorphic in $\lambda$ on each disc $D_x=\{|\lambda|\leq 1\}$ such
that $f=g+\bar g$ on $\p D_x$.  Such a $g$ can be obtained by the
integral formula \be{gdef} g(x,\lambda)=\frac1{2\pi
  i}\oint_{|\lambda'|=1} f(x^\alpha y^{\dot\alpha}+\lambda' \hat
x^\alpha \hat y^{\dot\alpha})\frac{\d \lambda'}{(\lambda'- \lambda)}
\, .  \ee Clearly $g$ is unique up to the addition of an imaginary
constant which, in the formula above, has been chosen so that
$g_0:=g(x,0)=\phi(x)$.

The key step in finding the ASDYM field from the ansatze is to find
the matrix functions $G(x,\lambda)$ holomorphic in $|\lambda|\leq 1$
that satisfy $GHG^*=I$ on $|\lambda|=1$ for the given
$$H=\begin{pmatrix}2Q^{-1}\cosh f&\e^{-f}
\\\e^{-f}&Q\e^{-f}\end{pmatrix}\, .$$  
If we set 
$$
G=\begin{pmatrix}a&b\\c&d\end{pmatrix}
$$
we should take $a,c$ to be sections of $\Oc(1)$ and $b,d$ of
$\Oc(-1)$ that are holomorphic over $D_x=\{|\lambda|\leq 1\}$ such
that
$$
a\e^f=\bar c+Q\bar d\, , \qquad \mbox{ and }\qquad (a+bQ)\e^{-f}=-\bar c\, .
$$
These equations can be
solved directly by expressing $f$ in terms of $g$ as above so that 
$$
(a+bQ)\e^{-g}=-\e^{\bar g}\bar c\, , \qquad \mbox{ and }\qquad
a\e^g=(\bar c+Q\bar d)\e^{-\bar g}\, .
$$
These expressions therefore determine sections of $\Oc(1)$ over $\p
D_x$ that extend over $D_x$ holomorphically and whose complex
conjugates do also.  By an application of an extension of Liouville's
theorem, they must therefore be the restriction of global sections
$\beta^A \lambda_A $ and $\alpha^A\lambda_A$ of $\Oc(1)$ where
$\alpha^A$ and $\beta^A$ are independent of $\lambda^A$.\footnote{The
  standard argument is that if both a function $g$ on $\p D_x$ and its
  complex conjugate $\bar g$ extend holomorphically over $D_x$, then
  its real and imaginary parts extend holomorphically over $D_x$ also
  and must therefore be constant.  Alternatively, a holomophic
  function $f$ on $D_x$ that is real on $\p D_x$ is constant since it
  can be extended to a bounded holomorphic function $\overline{
    f(1/\bar \lambda)}$ on the complex plain by inversion and
  continuity at $\p D_x$.  Here, the complex conjugates simply extend
  holomorphically over $D_x$ as $\bar \beta^A\lambda_A$ and $\bar
  \alpha^A\lambda_A$ since $\lambda_A=\bar \lambda_A$ on $\p D_x$).}

This gives for $a,b,c$ and $d$
$$
a=\e^{-g}  \alpha^A\lambda_A
\, , \; b=\frac{\e^g\beta^A\lambda_A-\e^{-g}
  \alpha^A\lambda_A}{Q}\, , \; c= -\e^{-g}\bar\beta^A\lambda_A\, ,
\; d=\frac{\e^g\bar \alpha^A\lambda_A +\e^{-g}\bar\beta^A\lambda_A}{Q}
$$
However, we require that at $Q=0$, $b$ and $d$ are regular and this
requires that the numerators of the fractions vanish there also.
Using $g(x,0)=\phi$, this
gives the relations
$$
\beta^0=\e^{-2\phi}\alpha^0\, , \quad \mbox{ and }\quad
\bar\beta^1=-\e^{2\phi}\bar\alpha^1 \, .
$$
These determine $\beta^A$ in terms of
$\alpha^A$ and reduce the unit determinant condition to
$$
1=ad-bc= (1+\e^{-4\phi})\alpha^0\bar\alpha^0 +
(1+\e^{4\phi})\alpha^1\bar\alpha^1 \, .
$$
The $J$-matrix is $J=G^{-1}(x,0)G^{*-1}(x,0)$ and calculation yields
that at $\lambda=0$:
$$
a=\e^{-\phi}\alpha^0\lambda_0\, , \quad b=
-\e^{-g}\frac{\alpha^1(1+\e^{4\phi})-2\alpha^0g'_0}{\lambda_0}\, ,$$
$$
c=\e^{\phi}\bar\alpha^1\lambda_0 \, , \quad d=\e^\phi\frac{2\bar\alpha^1
g'_0 + \bar \alpha_0(1+\e^{-4\phi})}{\lambda_0} 
$$
where $g'_0=\d g/\d\lambda |_{\lambda=0}$.  This gives 
$$
J=\begin{pmatrix}\frac{2(\cosh^22\phi+g_0'\bar g'_0)}{\lambda_0^2\cosh
    2\phi}  & \frac{-g'_0}{\cosh 2\phi}\\ \frac{-\bar g'_0}{\cosh
    2\phi} &\frac{\lambda_0^2}{2\cosh 2\phi}\end{pmatrix}\, .
$$
In this formula we note the appearance of $\lambda_0$ which is a
coordinate up the fibre of $\Oc(-1,-1)$.  A section $(s_0,s_1)$ of
$\Oc(1,1)\oplus\Oc(-1,-1)$ is here being understood as being represented
concretely by the homogeneous functions $(s_0/\lambda_0,s_1\lambda_1)$
and it is on expressions of this form that $J$ provides a hermitian
metric.

We also note the appearance of the function $g_0'(x)$.  This from
equation (\ref{gdef}) can be expressed as 
\be{gprime}
g'(x,0)=\frac1{2\pi i}\oint_{|\lambda'|=1} f(x^\alpha
y^{\dot\alpha}+\lambda' \hat 
x^\alpha \hat y^{\dot\alpha})\frac{\d 
\lambda'}{\lambda^2} \, .
\ee
This can be
obtained from $\phi$ as follows.  Recall that the Lax pair on the spin
bundle can be represented in this context by $L_0=\lambda\eth_x -
\bar\eth_y$ and $L_1=\lambda \eth_y-\bar\eth_x$.  Then, $L_0f=L_1f=0$
so that differentiation of equations (\ref{phi},\ref{gprime}) gives
$$
\bar\eth_x g'_0=\eth_y\phi\, , \quad \mbox{ and } \bar\eth_y g'_0=
\eth_x \phi\, .
$$
Given $\phi$ satisfying the wave equation, these equations are
integrable and can be solved for $g'_0$ in terms of $\phi$.

As a final comment, we note that the wave equation on $\phi$ in this
context is simply
$$
\Delta_x \phi=\Delta_y \phi$$
where $\Delta_x$ and $\Delta_y$ are the round sphere Laplacians on the
$x^A$ and $y^{A'}$ spheres respectively.  The equation therefore
clearly has separable solutions given as the product of spherical
harmonics $\phi=Y_{lm}(x^\alpha)Y_{lm'}(y^{\dot\alpha})$ on each
factor.  

\section*{References}
%\begin{thebibliography}

\smallskip\noindent 
Ablowitz, M. and Clarkson, P.A.  (1991) Solitons, nonlinear evolution
equations and inverse scattering, LMS lecture notes series 149, CUP.

\smallskip\noindent
Atiyah, M.F. (1979) Geometry of Yang-Mills fields, Accademia Nazionale
dei Lincei Scuola Normale Superiore, Lezione Fermiane, Pisa.

\smallskip\noindent
Atiyah, M.F., Hitchin, N., and Singer, I. (1978) Self-duality in
four-dimensional Riemannian geometry, {\it Proc. Roy. Soc. Lond.},
{\bf A 362}, 425-61.

\smallskip\noindent
Atiyah, M.F., \& Rees, E.G. (1976) Vector bundles on projective
3-space, Inventiones Mathematicae, {\bf 35}, 131-53.

\smallskip\noindent Bailey, T.N. (1985) Twistors and fields with
sources on worldlines, Proc. Roy. Soc. Lond., {\bf A397}, 143-55.

\smallskip\noindent
Bailey, T.N., Eastwood, M.G., Gover, R., and Mason, L.J. (1999) The
Funk transform as a Penrose transform, Math. Proc. Camb. Phil. Soc.,
{\bf 125}, no. 1. p67--81. 

\smallskip\noindent
Bailey, T.N., Eastwood, M.G., Gover, R., and Mason, L.J. (2003)
Complex analysis and the Funk transform, J. Korean Math Soc., {bf 40},
no. 4, 577-593.

\smallskip\noindent
Bailey, T.N., Eastwood (2001) Twistor results for integral transforms,
in {\em Radon transforms and tomography (South Hadley, MA, 2000)},
Contemp. Math. {\bf 278}, p77--86.

\smallskip\noindent
Belavin, A.A.,  and Zakharov, V.E. (1978) Yang-Mills equations as
inverse scattering problem, Phys. Lett., {\bf 73B}, 53-7. 

\smallskip\noindent
Berkovits, N. and Witten, E. (2004) Conformal gravity in
twistor-string theory, hep-th/0406051, see also hep-th/0402045 and
hep-th/0403187. 

\smallskip\noindent 
Chalmers, G. and Siegel, W. (1996) Self-dual sector of QCD amplitudes,
Phys. Rev. D, {\bf 54}, no. 12, 7628-33.

\smallskip\noindent 
Dunajski, M., (2002) Anti-self-dual 4-manifolds with a parallel real
spinor. {\em Proc.\ Roy. Soc. Lond.\ A}, {\bf 458}, no.\ 2021, p1205-22.

\smallskip\noindent 
Faddeev, L.D. and Takhtajan, L.A. (1987) Hamiltonian methods in the
theory of solitons, Springer-Verlag, Berlin-Heidelberg-New York.

\smallskip\noindent 
John, F. (1938) The ultrahyperbolic differential equation with four
independent variables, {\em Duke Math. J.}, {\bf 4}, 300-322,
reprinted in {\em 75 years of the Radon transform (Vienna 1992)},
Conf. Proc. Lecture Notes Math. Phys., IV, p301-323, International Press.

\smallskip\noindent
Gohberg, I.C., and  Krein, M.G. (1958) Systems of integral equations
on the half-line with kernels depending on the differences of the
arguments.  {\em Uspekhi Mat. Nauk},  {\bf 13}, 3-72. (Russian)

\smallskip\noindent
Guillemin, V.\ \& Sternberg, S. (1986) An ultra-hyperbolic analogue of
the Robinson-Kerr theorem. {\em Lett.\ Math. Phys.}, no.\ 1, 1--6.

\smallskip\noindent
Kotecha, V., Ward, R.S. (2001) Integrable Yang-Mills-Higgs equations
in three-dimensional de Sitter space-time, {\em J. Math.  Phys.}, {\bf
  42}, No. 3, 1018-1025.  

\smallskip\noindent
LeBrun and Mason (2002) Zoll manifolds and complex surfaces, {\em
  J. Diff. Geom.}, {\bf 61}, 453-535.

\smallskip\noindent
LeBrun and Mason (2005) Nonlinear Gravitons, Null Geodesics and
Holomorphic Discs, math.DG/0504582.

%\smallskip\noindent
%Lechtenfeld, O., and Popov, A. (2004) Supertwistors and Cubic string
%field theory for open N=2 strings, hep-th/0406179

\smallskip\noindent
Lerner, D.E. (1992) The linear system for self-dual gauge fields in a
space-time of signature 0, {\em J. Geom. Phys}, {\bf 8}, 211-9.

\smallskip\noindent
Manakov, S.V., \& Zakharov, V.E.\ (1981) Three dimensional model of
relativistic-invariant field theory, integrable by the inverse
scattering transform, {\em Lett. Math. Phys.}, {\bf 5}, 247-253. 

\smallskip\noindent 
Hughston, L.P., \& Mason, L.J. (1990) Further Advances in Twistor
Theory, Vol. I: The Penrose tranform and its applications, Pitman
Res. notes in Math. Ser., {\bf 231}, Longman.

\smallskip\noindent 
Mason, L.J. (1995)  Global solutions of the self-duality equations in
split signature, in {\em Further Advances In twistor Theory, Vol II:
Integrable Systems, conformal geometry and gravitation},  eds
L.J.Mason, L.P.Hughston \& P.Kobak, Pitman research Notes in Maths
Series, {\bf 232}, Longman Scientific and Technical / Wiley.

\smallskip\noindent
Mason, L.J. \& Woodhouse, N.M.J. (1996) {\it Integrability, self-duality
and twistor theory}, OUP.

\smallskip\noindent
Novikov, R.\ (2002) On determination of a gauge field on $\R^d$ from
its non-abelian Radon transform along oriented straight lines,
J.Inst. Math. Jussieu, {\bf 1}, 4, 559-629.

\smallskip\noindent
Penrose, R. (1963 reprinted 1980) Null hypersurface initial data for
classical fields of arbitrary spin and for general relativity, in {\em
  Aerospace Research Laboratories Report 63-56} (P.G.Bergmann)
reprinted (1980) in {\em Gen. Rel. Grav.}, {\bf 12}, 225-64.

\smallskip\noindent Penrose, R. (1976) Nonlinear gravitons and curved
twistor theory, Gen. Rel. Grav. {\bf 7}, 31-52.

\smallskip\noindent Penrose, R., \& Rindler, W.\ (1984 \& 1986)
{\em Spinors \& Space-times Vol. I \& II}, CUP.

%\smallskip\noindent Popov. A., and S\"amann, C. (2004) On
%supertwistors, the Penrose transform and ${\mathcal N}=4$ super
%Yang-Mills theory, hep-th/0405123.

\smallskip\noindent
Rendall, A. (1992) The characteristic initial value problem for the
Einstein equations, {\em Nonlinear hyperbolic equations and field
  theory (Lake Como, 1990)}, p154--163, Pitman Res. Notes Maths
Ser., {\bf 253}.  

\smallskip\noindent
Sparling, G.A.J. (1998) Inversion for the Radon line transform in
higher dimensions, Phil. Trans. Roy. Soc. Ser. A, {\bf 356}, no. 1749,
p3041--3086.

%\smallskip\noindent
%Villarroel, J. (1991) Yang-Mills equations and the inverse scattering
%transform, J.Phys. A, {\bf 24}, no. 15, 3587-92.

\smallskip\noindent
Ward, R.S.\ (1977) On self-dual gauge fields, Phys. Lett., {\bf 61A}, 81-2. 

\smallskip\noindent
Ward, R.S., (1979) Massless fields from twistor functions, \S2.4 in {\em
  Advances in Twistor Theory}, eds Hughston \& Ward, Pitman Research
Notes in Math., {\bf 37}, Pitman.

\smallskip\noindent
Ward, R.S. (1995) Nontrivial scattering of localized solitons
in a $(2+1)$-dimensional integrable system.  {\em Phys. Lett. A}, {\bf
  208}, no.\ 3, 203-8.

\smallskip\noindent
Ward, R.S. (1999) Two integrable systems related to hyperbolic
monopoles, in {\em Sir Michael Atiyah: A great mathematician of the
  twentieth century, Asian J.Math.}, {\bf 3}, no. 1, 325-332.

\smallskip\noindent
Ward, R.S., \& Wells, R.O. (1990) Twistor geometry and field theory, CUP.

\smallskip\noindent
Witten, E. (2003) Perturbative gauge theory as a string theory in
twistor space, hep-th/0312171.

\smallskip\noindent
Woodhouse, N.M.J. (1992) Contour integrals for the ultrahyperbolic
wave equation, Proc.\ Roy.\ Soc.\ London, {\bf A438}, 197-206.

\noindent\noindent
\noindent
\smallskip\noindent \smallskip\noindent \smallskip\noindent \smallskip\noindent
%\end{thebibliography}

\end{document}